\documentclass[5p]{elsarticle}
\makeatletter
\def\ps@pprintTitle{%
 \let\@oddhead\@empty
 \let\@evenhead\@empty
 \def\@oddfoot{}%
 \let\@evenfoot\@oddfoot}
\makeatother
\usepackage[utf8]{inputenc}
\usepackage[OT1]{fontenc}
\usepackage[english]{babel}

\usepackage{bm}
\usepackage{dsfont}
\usepackage{csquotes}

\usepackage{graphicx}
\usepackage{url}

\usepackage[table]{xcolor}
\usepackage{tabularx}
\usepackage{booktabs}  
\usepackage{multirow}

\usepackage[textsize=tiny]{todonotes}

\usepackage{xspace}
\usepackage{nicefrac}
\usepackage[super]{nth}

\usepackage{amsmath}
\usepackage{amsfonts} 
\usepackage{amssymb}
\usepackage{mathrsfs}
\usepackage{mathtools} 
\mathtoolsset{showonlyrefs}

\usepackage{amsthm}
\usepackage{enumitem}
\setlist[itemize]{itemsep=.1ex, parsep=.1ex, topsep=.1ex,
   leftmargin=1.5ex, labelsep=.5ex}
\setlist[description]{font=\bfseries\sffamily, topsep=.1ex, parsep=.2ex}
\setlist[enumerate]{font=\bfseries\sffamily, topsep=.2ex,
   parsep=.15ex, leftmargin=4ex}

\definecolor{deep_blue}{rgb}{0,.2,.5}
\definecolor{dark_blue}{rgb}{0,.15,.5}

\usepackage{hyperref}
\hypersetup{pdftex,  
  breaklinks=true,
  colorlinks=true,
  linkcolor=dark_blue,
  citecolor=deep_blue,
}

\makeatletter
\addto\extrasenglish{%
\renewcommand{\sectionautorefname}{\S\@gobble}%
\renewcommand{\subsectionautorefname}{\S\@gobble}%
\renewcommand{\subsubsectionautorefname}{\S\@gobble}%
}
\makeatother

\usepackage{tikz}
\usetikzlibrary{spy}

\usepackage[para]{threeparttable}

\newcommand{\RR}{\mathbb{R}}

\newcommand{\A}{\bm{A}}

\newcommand{\D}{\bm{D}}

\newcommand{\R}{\bm{R}}

\newcommand{\X}{\bm{X}}
\newcommand{\Y}{\bm{Y}}

\renewcommand{\d}{\bm{d}}

\newcommand{\x}{\bm{x}}
\newcommand{\y}{\bm{y}}
\newcommand{\z}{\bm{z}}

\newcommand{\bal}{\bm\alpha}
\newcommand{\bbe}{\bm\beta}

\renewcommand{\phi}{\varphi}

\newcommand{\foralls}{\forall \,}

\renewcommand{\leq}{\leqslant}
\renewcommand{\geq}{\geqslant}

\renewcommand{\epsilon}{\varepsilon}
\renewcommand{\imath}{\mathrm{i}}

\DeclareMathOperator*{\argmin}{argmin}

\newlength{\restsubwidth}
\newlength{\restsubheight}
\newlength{\restsubmoreheight}
\newcommand{\rest}[2]{%
        \settowidth{\restsubwidth}{\ensuremath{#2}}
        \settoheight{\restsubheight}{\ensuremath{{}_{#2}}}
        \ensuremath{{#1\hskip 0.5pt}_{\vrule\kern2pt\parbox[b][%
        4pt][b]{\the\restsubwidth}{%
                        \ensuremath{{}_{#2}}}}}
        }

\definecolor{dred}{rgb}{0.8,0,0}
\definecolor{dgreen}{rgb}{0,0.8,0}
\definecolor{dblue}{rgb}{0,0,0.8}
\definecolor{dpurple}{rgb}{0.8,0,0.8}

\biboptions{authoryear}

\begin{document}

\title{Fine-grain atlases of functional modes for fMRI analysis}

\author[parietal]{Kamalaker Dadi}
\author[parietal]{Gaël Varoquaux}
\author[parietal]{Antonia Machlouzarides-Shalit}
\author[stanford]{Krzysztof J. Gorgolewski}
\author[parietal]{Demian Wassermann}
\author[parietal]{Bertrand Thirion}
\author[parietal,ens]{Arthur Mensch}
\address[parietal]{Inria, CEA, Université Paris-Saclay, Palaiseau, 91120, France}
\address[ens]{ENS, DMA, 45 rue d'Ulm, 75005 Paris}
\address[stanford]{Department of Psychology, Stanford University, California,
USA}

\begin{abstract}
Population imaging markedly increased the size of
functional-imaging datasets, shedding new light on the neural basis
of inter-individual differences. 
Analyzing these large data entails new scalability challenges, 
computational and
statistical.
For this reason, brain images are typically summarized in a few signals,
for instance reducing voxel-level measures with brain
atlases or functional modes.
A good choice of the corresponding brain networks is important, as
most data analyses start from these reduced signals.
We contribute finely-resolved atlases of functional modes,
 comprising from 64 to 1024 networks.
These dictionaries of functional modes (DiFuMo) are trained on millions of
fMRI functional brain volumes of total size 2.4TB, spanned over 27
studies and many research groups.
We demonstrate the benefits of extracting reduced signals on our fine-grain
atlases for many classic functional data analysis pipelines: stimuli decoding from
12,334 brain responses, standard GLM analysis of fMRI across sessions and
individuals, extraction of resting-state functional-connectomes biomarkers for
2,500 individuals, data compression and meta-analysis over more than 15,000
statistical maps.
In each of these analysis scenarii,
we compare the performance of our functional atlases with that of other popular
references, and to a simple voxel-level analysis.
%
Results highlight the importance of using high-dimensional ``soft'' functional
atlases, to represent
 and analyse brain activity while capturing its functional gradients.
Analyses on high-dimensional modes achieve similar statistical performance as at
 the voxel level, but with much reduced computational cost and higher interpretability.
In addition to making them available, we provide meaningful names for these modes, based on
their anatomical location. It will facilitate reporting of results.
%
%

\end{abstract}


\maketitle

\sloppy

\section{Introduction}
%
Population imaging has been bringing in terabytes of high-resolution functional
brain images, uncovering the neural basis of individual differences
\citep{elliott2008}.
%
%
While these great volumes of data enable fitting richer statistical
models, they also entail massive data storage
 \citep{poldrack2013openfmri, gorgolewski2017} and challenging
high-dimensional data analysis.
A popular approach to facilitate data handling is to work with image-derived
phenotypes (IDPs), i.e. low-dimensional signals that summarize the
information in the images while keeping meaningful representations of the
brain \citep{miller2016}.

While brain atlases originated in characterizing the brain's
microstructure \citep{brodmann1909vergleichende}, today they are widely
used to study functional
connectomes \citep{sporns2005,varoquaux2013connectomes} and for data
reduction in functional imaging \citep{thirion2006,craddock2012}.
For these applications, the choice of brain regions conditions the signal
captured in the data analysis.
To define regions well suited to brain-imaging endeavors, there is great
progress in building atlases from the neuroimaging data itself
\citep{eickhoff2018}.
Yet, most functional atlases describe the brain as parcellations,
locally-uniform functional units, and thus do not represent well
functional gradients \citep{huntenburg2018large}.

For functional imaging, brain structures delineated by an atlas
should capture the main features of the functional signal, e.g. the
functional networks \citep{smith2011}.
In a nutshell, there are two approaches to define well-suited structures.
These can strive to select \emph{homogenous} neural
populations, typically via clustering approaches
\citep{goutte1999clustering,bellec2010,craddock2012,thirion2014,schaefer2018}. They can
also be defined via continuous \emph{modes} that map intrinsic
brain functional networks \citep{damoiseaux2006,varoquaux2011,harrison2015}.
These functional modes have been shown to capture well
functional connectivity, with techniques such as  
Independent Component Analysis \citep{kiviniemi2009, pervaiz2019} or
sparse
dictionary learning \citep{mensch2016, dadi2019}.
%

%
%

High-resolution atlases can give a fine-grained division of the brain and 
capture more functionally-specific
regions and rich descriptions of brain activity \citep{schaefer2018}. Yet, there is to date no highly-resolved set of
``soft'' functional modes available, presumably
because increasing the dimensionality raises significant 
computational and statistical challenges  
\citep{mensch2016icml, pervaiz2019}.
In this paper, we address this need with high-order
dictionaries of functional modes (DiFuMo) extracted at a large scale
both in terms of
data size (3 million volumes of total data size 2.4TB) and resolution (up to 1024 modes).
For this, we leverage the wealth of openly-available functional images
\citep{poldrack2013openfmri} and efficient dictionary-learning
algorithms to fit on large data.
This is unlike ICA which is hard to use for a high number of modes \citep{pervaiz2019}.

\paragraph{Contributions}
We provide Dictionaries of Functional
Modes\footnote{\url{https://parietal-inria.github.io/DiFuMo}} ``\emph{DiFuMo}'' that can
serve as atlases to extract functional signals, e.g. provide IDPs, with different
dimensionalities (64, 128, 256, 512, and 1024).
These modes are optimized to represent BOLD data well, over a
wide range of experimental conditions. They are more finely-resolved
than existing brain decompositions with continuous networks.
By providing validated fine functional atlases, our goal is to streamline fMRI 
analysis with reduced representations, to facilitate large-cohort and
inter-studies work.
Through thorough benchmarking over classic data analysis tasks, we show that these
modes gives IDPs that ground better analysis of functional images.
Finally, we provide a meaningful label to each mode, summarizing
its anatomical location, to facilitate reporting of results.

\section{Methods: data-driven fine-grain functional modes}

We describe in this section the models and methods underlying our definition of
brain structures to extract IDPs.

\subsection{Context: Image Derived Phenotypes}

While analysis of brain images has been pioneered at the
voxel level \citep{friston1995}, image-derived phenotypes (IDP) are 
increasingly used in the context of population imaging.
Trading voxel-level signals for IDPs has several motivations. First and
foremost, it greatly facilitates the analysis on large cohorts: the data are smaller, easier
to share, requiring less disk storage, computer memory, and computing
power to analyze. It can also come with statistical benefits. For
instance, in standard analysis of task responses, e.g. in mass-univariate brain mapping, 
the statistical power of hypothesis test at the voxel level is limited by
multiple comparisons~\citep{friston1995}, while working at the level of
IDPs mitigates this problem~\citep{thirion2006}.
For predictive modeling, e.g. in multi-variate
decoding~\citep{mouraomiranda2005}, the high-dimensionality of the signals 
is a challenge to learning models that generalize well---a phenomenon known
as the curse of dimensionality in machine learning~\citep{hastie2009elements}. Finally, for functional
connectomes, working at voxel-level is computationally and statistically
intractable as it entails modeling billions of connections. The standard approach is therefore to average signals on regions or networks
\citep{varoquaux2013connectomes}.

Functional neuroimaging is currently largely dependent on
neuroanatomy for mapping function to structure
\citep{destrieux2010, devlin2007}.
Some anatomical structures support well a direct mapping to specific
functions \citep{brett2002, rademacher1993}, e.g. the primary
visual areas.
Yet other functional units
are not simply defined from anatomical features, for instance in
high-level regions such as the default mode, which is
defined from functional data \citep{leech2010,greicius2003}.

\begin{figure}[b!]
    \centerline{%
        \includegraphics[width=0.9\linewidth]{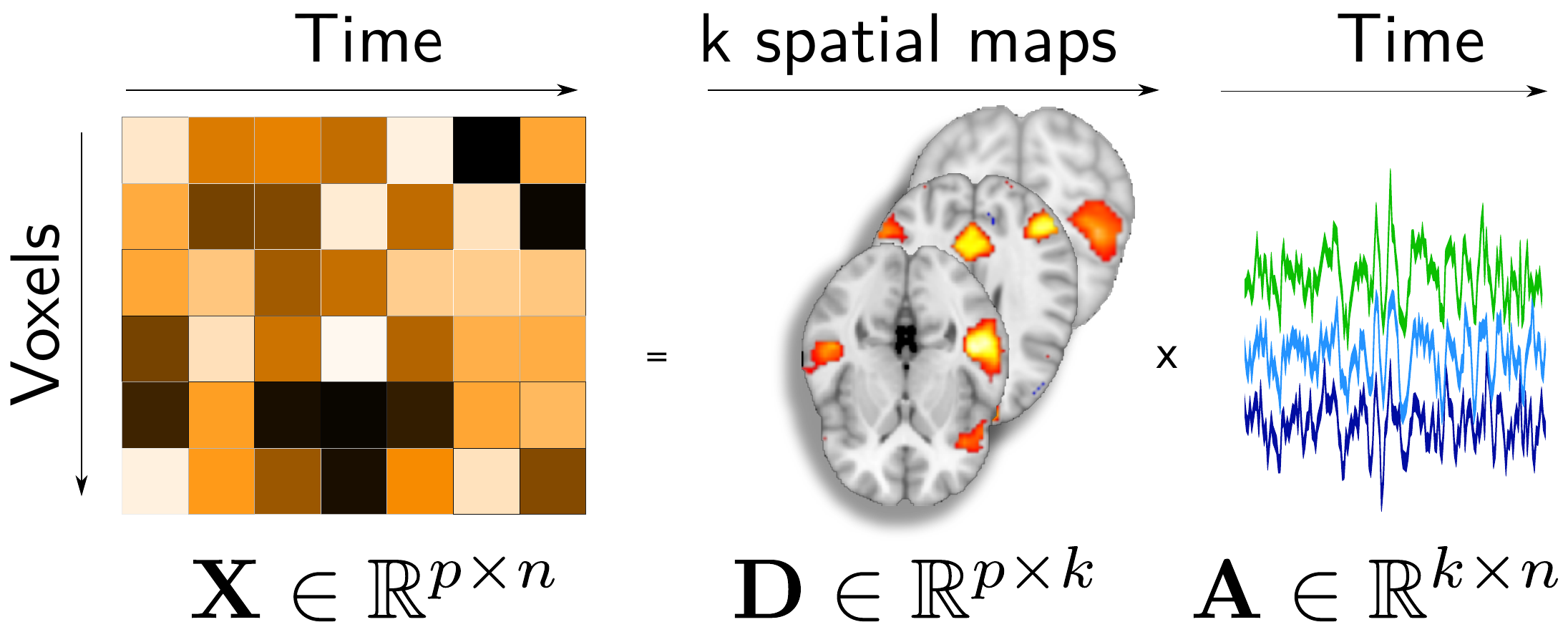}%
    }%
    \caption{\textbf{Linear decomposition model of fMRI time-series for estimating
        brain networks}: The fMRI time series $\textbf{X}$ are factorized into a product
        of two matrices, $\D$ wich contain spatial modes and $\textbf{A}$
        temporal loadings of each mode. $p$ - number of features, $n$ -
        number of volumes in fMRI image, $k$ - number of dictionaries.
    \label{fig:matrix_factorization}
    }
\end{figure}

\subsection{Methods for data-driven functional atlases}

Compared to anatomical atlases, defining regions from the functional
signal can lead to a better explanation of behavioral
outcomes \citep{dadi2019}, as
they capture the functional structure of the brain.
Clustering of fMRI timeseries has been heavily used to define brain
parcellations \citep{goutte1999clustering}, or for data reduction in predictive models
\citep{michel2012supervisedclustering}. Reference functional brain
parcellations have been defined with various clustering algorithms on
resting-state fMRI \citep{bellec2010,yeo2011,craddock2012}.
Another class of approaches seeks modes of brain activity, decomposing
the signal as a product of spatial maps and corresponding
time-series (\autoref{fig:matrix_factorization}). The most popular model
in neuroimaging
is independent component analysis  \citep[ICA,][]{hyvarinen2000}, which 
optimizes spatial independence between extracted maps.
It has been extensively used to define resting-state networks
\citep{kiviniemi2003,beckmann2005,calhoun2001a} and implicitly outlines
soft parcellations of the brain at high order
\citep{kiviniemi2009,varoquaux2010}.
ICA-defined networks are used to extract the official IDPs of UK BioBank,
the largest brain-imaging cohort to date; these have been
shown to relate to behavior \citep{miller2016}.

We rely on another decomposition model, dictionary learning
\citep{olshausen1997}, that enforces sparsity and non-negativity
instead of independence on the spatial maps. While less popular than ICA in
neuroimaging,
sparsity brings the benefit of segmenting well functional
regions on a zeroed-out background \citep{lee2010data,varoquaux2011}.
For our purposes, an important aspect of sparse models is that they 
have computationally-scalable formulations even with high model
order and on large datasets \citep{mensch2016icml,mensch2018ieee}.
Functional modes defined from 
sparse dictionary learning have been used to predict Autism 
Spectrum Disorder \citep{abraham2017deriving}, or mental
processes \citep{mensch2017nips}.

\paragraph{Rest and task fMRI} Most functional brain atlases have
been extracted from rest fMRI
\citep{bellec2010, power2011atlas, craddock2012, yeo2011, miller2016,
schaefer2018}. Brain networks can also be extracted from task fMRI data
\citep{calhoun2008, lee2010data}, and segment a similar intrinsic
large-scale structure \citep{smith2009}. In our work, we build
functional modes from datasets with different experimental
conditions, including task and rest. Our goal is to be as general as
possible and capture information from different protocols. Indeed, 
defining networks on task fMRI can help representing these brain images
and predicting
the corresponding psychological conditions \citep{duff2012}.
%


\subsection{DiFuMo extraction: model and data}
\label{sec:matrix_factorization}

We consider BOLD time-series from fMRI volumes,
resampled and registered to the MNI template. After temporal concatenation, those
form a large matrix $\X \in \RR^{p \times n}$, where $p$ is the number of voxels of
the images (around $2\cdot10^5$), and $n$ is the number of brain images, of the order
of $10^6$ in the following.
To extract DiFuMos,
each brain volume is modeled as the linear combination
of $k$ spatial functional networks, assembled in a dictionary matrix
$\D \in \RR^{p \times k}$.
We thus assume that $\X$ approximately factorizes as $\D \A$,
where the matrix $\A \in \RR^{k \times n}$ holds in every column the loadings $\bal_i$ necessary to reconstruct the brain image $\x_i$ from the networks $\D$.
The \textit{dictionary} $\D$ is to be \textit{learned} from data.
For this, we rely on \emph{Stochastic Online Matrix
Factorization}\footnote{Available at:
\url{https://arthurmensch.github.io/modl/}} \citep[SOMF]{mensch2018ieee}, that is computationally tractable for 
matrices large in both directions, as with high-resolution large-scale fMRI 
data. SOMF solves the constrained $\ell_2$ reconstruction problem
\begin{equation}
    \min_{
        \substack{\D \in \RR^{p \times k}, \A \in \RR^{k \times n}\\
        \D \geq 0, \foralls j \in [k],\,\Vert \d_j \Vert_1 \leq 1
    }
    }
    \Vert \X - \D \A \Vert_F^2 + \lambda \Vert \A \Vert_F^2,
\end{equation}
where $\lambda$ is a regularization parameter that controls the sparsity
of the dictionary $\D$, via the $\ell_1$ and positivity constraints. Encouraging
sparsity in spatial maps is key to obtaining well-localized maps
that outline few brain regions. The parameter $\lambda$ is chosen so that the union of
all maps approximately covers the whole brain.

\begin{figure*}[t]
    \includegraphics[width=1.\linewidth]{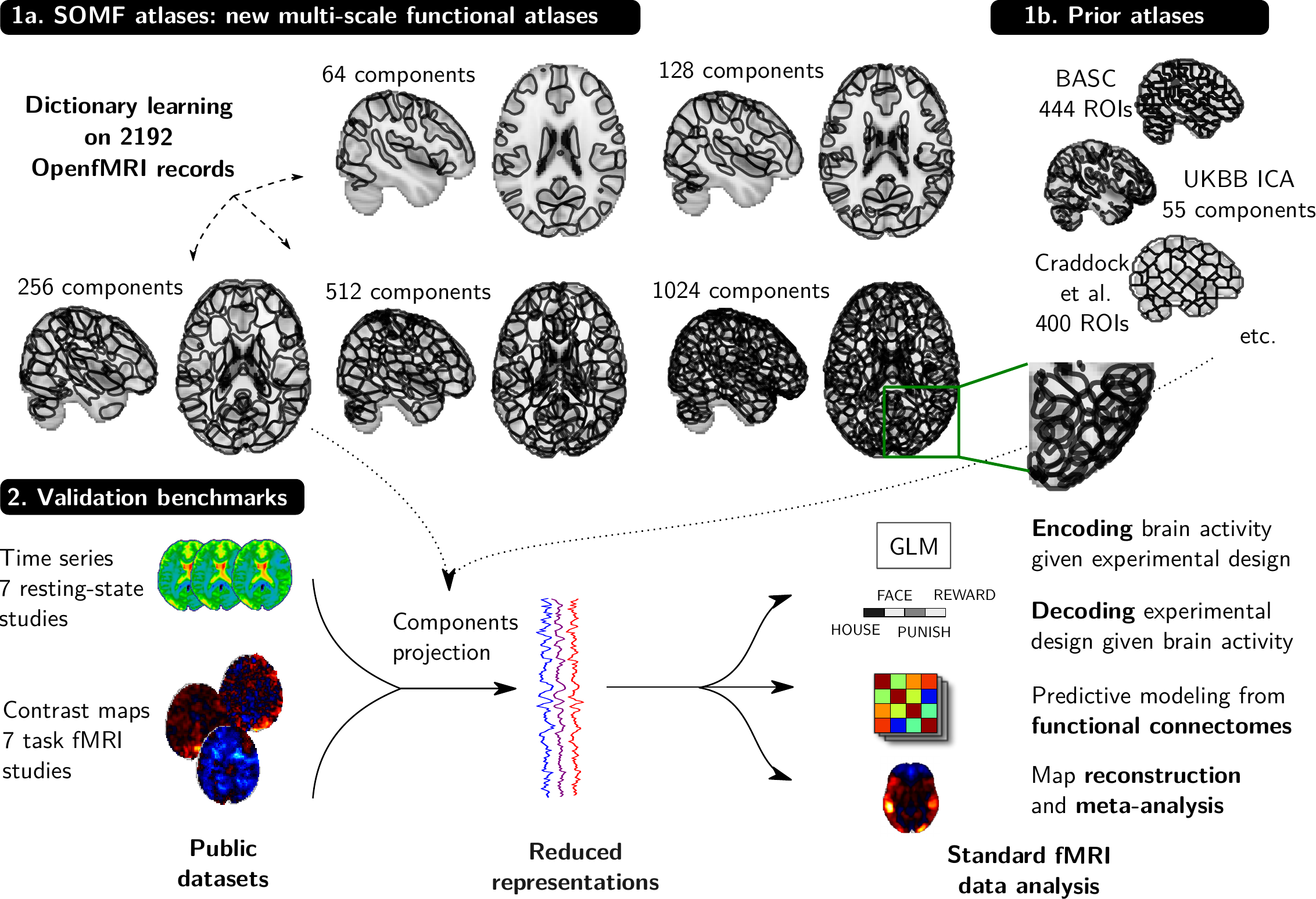}%
    \caption{\textbf{Schema of DiFuMo atlases and their usage in 
	typical fMRI analyses.} DiFuMo atlases are extracted from a massive concatenation of
	BOLD time-series across fMRI studies, using a sparsity inducing matrix factorization
	algorithm.
      We compute the DiFuMo atlases at different resolutions, up to 1024 components.
    We assess our atlases in 4 benchmarks
	that measure suitability to classic fMRI analyses.
      Those are performed on reduced and non-reduced data, with different atlas
	sizes and a comparison between atlases.
    The easiest way to view and download DiFuMo atlases is via the online
    interactive visualizations:
    \href{https://parietal-inria.github.io/DiFuMo}{parietal-inria.github.io/DiFuMo}.
	\label{fig:abstract}
    }
\end{figure*}

\paragraph{Input fMRI data} We build the input data matrix $\X$ with
BOLD time-series from 25 different task-based fMRI studies and 2 resting
state studies, adding up to $2\,192$ functional MRI recording sessions. We gather data
from OpenNeuro \citep{gorgolewski2017}
--\autoref{tab:task_openneuro} lists the corresponding studies 
while
\autoref{tab:data_acquisition_somf_learner} gives their
data-acquisition parameters.

We use \textit{fMRIprep} \citep{esteban2019fmriprep} for minimal preprocessing:
brain extraction giving as a
reference to correct for head-motion \citep{jenkinson2002}, and
co-registration to anatomy \citep{greve2009}.
All the fMRI images are transformed to MNI template space.
We then use
\textit{MRIQC} \citep{esteban_mriqc} for quality control.

\paragraph{Multi-dimensional DiFuMo atlases} We estimate dictionaries of
dimensionality $k \in \{64, 128, 256, 512, 1024\}$. This is useful as the
optimal dimensionality for extracting IDPs often depends on the downstream data
analysis task.
The obtained functional modes segment well-localized regions, as illustrated in
\autoref{fig:abstract}.

\subsection{Extracting signal on functional modes}
\label{sec:reduced_representations}

The functional modes take continuous values (we refer to them as \textit{soft}) and can have some overlap
--though in practice this overlap is small. As a consequence, signal extraction calls for
more than averaging on regions.
The natural formulation is that the extracted signals (the IDPs)
should best approximate the brain image $\x \in \RR^p$ as a linear combination $\bal \in \RR^k$ of
the set of modes in the dictionary $\D \in \RR^{p \times k}$.
This is solved by linear regression:
\begin{equation}\label{eq:reduction}
    \bal = \argmin_{\bal \in \RR^{k}} \Vert \x - \D \bal \Vert_2^2,\quad\text{i.e.}\quad \bal = \D^\dagger \x,
\end{equation}
where $\D^\dagger = (\D^\mathsf{T} \D)^{-1} \D^\mathsf{T}\in \RR^{k \times p}$ is the pseudo-inverse of~$\D$.
For atlases composed of non-overlapping regions, such as classic brain
parcellations---e.g. BASC
\citep{bellec2010} or normalized cuts \citep{craddock2012}---linear regression
simply amounts to averaging the images values in every cluster of $\D$.
For overlapping modes as the ones of DiFuMo or the ICA maps used in UKBB
\citep{miller2016}, the linear regression formulation caters for the
overlap and softness of the regions.

\subsection{Region names: relation to anatomical structures}

Relating IDPs to known brain structures facilitates interpretation and
discussion of results.
Though the DiFuMo atlases are defined from functional signal, we choose
to reference their regions by their anatomical location, as it is a common
and well-accepted terminology in neuroscience.
For each resolution, we match the modes with regions in references of
brain structure:
the Harvard-Oxford atlas \citep{desikan2006}, Destrieux atlas \citep{destrieux2010},
the MIST atlas \citep{urchs2019mist}, Johns Hopkins
University (JHU) atlas \citep{hua2008}, and the Dierdrichsen cerebellum atlas \citep{diedrichsen2009}. We name each mode from the anatomical
structure that it most overlaps with. When the overlap was
weak, a trained neuroanatomist (AMS) looked up the structure in 
standard classic anatomy references
\citep{henri1999, schmahmann1999, rademacher1992, michio1990, marco2012}.
\ref{app:difumo_labeling} gives
more details on the naming of the brain areas.

\section{Brain-image analysis on functional modes}
\label{sec:pipelines}%
\enlargethispage{1ex}

We use the reduced representations (IDPs) introduced above
for various functional-imaging analytic tasks:
standard mass-univariate analysis
of brain responses (\autoref{sec:glm-pipeline});
decoding of mental processes
from  brain activity (\autoref{sec:decoding-pipeline});
prediction of phenotypes from functional connectomes
(\autoref{sec:rsfmri-pipeline});
finally, we measure the quality of
signal reconstruction after the dimension reduction,
with an illustration on meta-analyses
(\autoref{sec:image-rec-pipeline}).

\begin{table*}[t]
\begin{threeparttable}\small%
\rowcolors{2}{gray!20}{white}
\begin{tabular}{lp{24.5ex}lrp{32ex}r}
	\!Name & Dimensionality & \hspace*{-2ex}\# subj.\!& \hspace*{-2ex}Soft\! & Extraction method &
\hspace*{-6ex}Reference\!\!\\
\midrule
\!BASC           & 64, 122, 197, 325, 444 & 43 & No & Hierarchical clustering &  \cite{bellec2010}\!\\
\!Craddock       & 200, 400 & 41 & No & Spectral clustering & \cite{craddock2012}\!\\
	\!FIND\tnote{a} & 90, 499 & 15 & Yes & ICA; Ward clustering & \hspace*{-5ex}\cite{shirer2012,altmann2015}\!\\
\!Gordon         & 333 & 120 & No & Local-gradient approach & \cite{gordonatlas2014}\!\\
\!UKBB ICA\!     & 21, 55    & 4100 & Yes & Selected ICA
components\tnote{b} & \cite{miller2016}\!\\
\!Schaefer       & 100, 200, 300, 400, 500, 600, 800, 1000 & 1489 & No & Gradient-weighted
                Markov Random Field (gwMRF)
               & \cite{schaefer2018}\!\\
\textbf{DiFuMo}\tnote{c} & 64, 128, 256, 512, \textbf{1024} & \textbf{2192} & \textbf{Yes} & \textbf{Sparse dictionary
learning} & \textbf{This paper}\\
\midrule
\end{tabular}
\begin{tablenotes}\footnotesize
  \item [a] \url{https://findlab.stanford.edu/functional_ROIs.html}
  \item [b] \url{https://www.fmrib.ox.ac.uk/ukbiobank/}
  \item [c] \url{https://parietal-inria.github.io/DiFuMo}
\end{tablenotes}
\end{threeparttable}
    \caption{Functional atlases that we benchmark; they define IDPs for brain-images.
analysis}%
    \label{tab:atlases}
\end{table*}

\subsection{Benchmarking several functional atlases}

To gauge the usefulness of the extracted IDPs, we compare 
each analysis pipeline across several functional atlases:
DiFuMo and reference atlases
are used to compute functional IDPs.
We use the same signal-extraction function \eqref{eq:reduction}, but
vary the spatial components $\D$.
As a baseline, we also perform the voxel-level analyses,
though it entail significantly larger computational costs.

\enlargethispage{2ex}%

We consider other functional atlases that are
multi-resolutions, accessible to download, and volumetric
(\autoref{tab:atlases}):
\textit{ICA maps} with $k \in \{21, 55\}$ components, extracted on
large-scale rs-fMRI from UKBB \citep{miller2016};
\textit{bootstrap analysis of stable clusters (BASC)}
built with hierarchical clustering on rs-fMRI, with various
number of clusters \citep{bellec2010};
spatially-constrained clustering on
rs-fMRI, with $k \in \{200, 400\}$ clusters \citep{craddock2012};
$k = 333$ cortical areas derived from 
rs-fMRI using a local gradient approach \citep{gordonatlas2014};
$k \in \{90, 499\}$ functional regions covering cortical
and subcortical gray matter with ICA and Ward clustering
(\cite{shirer2012},\linebreak\cite{altmann2015});
and brain parcellations derived with gradient-weighted Markov Random Field,
with resolutions
similar to ours \citep[][$k$ up to 1000]{schaefer2018}.

\subsection{Mapping brain response: standard task-fMRI analysis}\label{sec:glm-pipeline}

Standard analysis in task fMRI
relates psychological manipulations to brain activity separately for each voxel or region.
It models the BOLD signal as a linear combination of experimental conditions---the General Linear Model \citep[GLM,][]{friston1995}.
The BOLD signal forms a matrix $\Y \in \RR^{n \times p}$, where $p$ is the number of voxels.
With data reduction, we use as input the reduced signal
$\Y_{\text{red}} = \Y_{\text{voxel}} (\D^\dagger)^\top \in \RR^{n \times
k}$ (\autoref{eq:reduction}).
The GLM models $\Y$ or $\Y_{\text{red}}$ as $\Y = \X \bm \beta + \bm \epsilon$
where $\X \in \RR^{n \times q}$ is the design matrix formed by $q$ temporal regressors of interest or nuisance
and $\bm \epsilon$ is noise \citep{friston1998}. In our experiments, we use the 
\textit{Nistats} library\footnote{\url{https://nistats.github.io/}}.

With reduced input $\Y_{\text{red}}$, we obtain one signal per region, as $\bbe \in \RR^{q \times k}$.
The full $\beta$-maps can then be reconstructed by setting $\bbe_{rec} = \bbe \D^\top \in \RR^{q \times p}$.
We transform the reconstructed $\beta$-maps into
 \emph{z-maps} $\z \in \RR^{q \times p}$ using base contrasts, before
thresholding them with \citet[]{fdr_ben_hoch_1995} FDR correction
for multiple comparisons.
We then compare the $z$-maps obtained using voxels as input, and $z$-maps
using reduced input and reconstructed $\beta$-maps, using the \citet{dice} similarity coefficient.
We also perform an \emph{intra-subject} analysis 
detailed in \ref{sec:intra_subject_analysis}.

\enlargethispage{2ex}%

\paragraph{Data}
We consider the Rapid-Serial-Visual-Presentation (RSVP)
language task of Individual Brain Charting (IBC)
\citep[see][for experimental protocol and pre-processing]{pinho2018ibc}.
We model six experimental conditions: complex meaningful sentences, simple
meaningful sentences,
jabberwocky, list of words, lists of pseudowords, consonant strings.
$\beta$-maps are estimated for each subject using a fixed-effect model
over 3 out of the 6 subject's sessions.
We randomly select 3 sessions 10 times to estimate the variance of the Dice index across sessions.
As a baseline, we evaluate the mean and variance of the Dice index across $z$-maps when varying the sessions used in voxel-level GLM.

\subsection{Decoding experimental stimuli from brain responses}%
\label{sec:decoding-pipeline}
\emph{Decoding} predicts psychological conditions from
task-related $z$-maps \citep{haynes2006}.
The validity of a decoding model is evaluated on left-out data 
\citep[following][]{varoquaux2017assessing}, e.g.
left-out subjects for inter-subject decoding \citep{poldrack2009decoding}.
We use linear decoding models: ridge regression for continuous
target and Support Vector Machine \citep[SVC,][]{hastie2009elements}
for classification.
For each study, we separate sessions (for intra-subject decoding)
or subjects (for inter-subject decoding)
into randomly-chosen train and test folds 
(20 folds with $30\%$ test size), and measure the
test accuracy. 
We compare the performance of predictive models using the voxel-level $z$-maps or using the
data reduced with functional atlases.

\paragraph{Data}
We use 6 open-access task fMRI studies.
We perform \emph{inter-subject} decoding in
the \textbf{emotional} and sensitivity to \textbf{pain} experiences
from \cite{chang2015emotion}, and in three studies from HCP900 \citep{vanessen2012hcpdata}: \textbf{working memory},
\textbf{gambling} \citep{delgado2000gambling},
and \textbf{relational processing} \citep{smith2007relational}.
We perform \emph{intra-subject} decoding using
the several sessions of \textbf{left} and \textbf{right} button press responses in IBC \citep[ARCHI protocol,][]{pinel2007}.
The unthresholded $z$-maps
used in the decoding pipeline are either obtained from Neurovault
\citep{gorgolewski2015}, or computed with the GLM following \autoref{sec:glm-pipeline}.
Details are reported in \ref{sec:task_fMRI_data}.

\subsection{Predicting phenotype from functional connectomes}
\label{sec:rsfmri-pipeline}
Resting-state fMRI can be used to predict phenotypic traits
\citep{richiardi2010}.
For this, each subject is represented by a functional connectivity matrix 
that 
captures the correlation between brain signals at various locations.
Our
functional-connectome prediction pipeline comprises three steps:
\textbf{1)} we extract a 
reduced representation of the BOLD signal, projecting voxel-level data
onto a functional atlas as in \autoref{sec:glm-pipeline}; \textbf{2)}
we compute a \textit{functional connectome} from the reduced BOLD signals;
\textbf{3)} we use it as input to a linear model.
We compute a connectome from activations with
the \citet{ledoit2004} covariance estimator as
\citet{varoquaux2013connectomes, brier2015partial}.
We then derive
single-subject features from covariance matrices using their
tangent space parametrization \citep{varoquaux2010b,
barachant2013,pervaiz2019}.
Those are used to fit an $\ell_2$-penalized logistic regression for
classification and a ridge regression for continuous targets.
We assess predictive performance with 20 folds,
random splits of subjects in train and test sets, with $25\%$ test size.

\paragraph{Data}
We use 7 openly-accessible datasets with diverse phenotypic
targets, as summarized in \autoref{tab:rest}. We predict diagnostic
status for Alzheimer's disease on \textbf{ADNI} \citep{mueller2005},
PTSD on \textbf{ADNIDOD}; Autism Spetrum Disorder on
\textbf{ABIDE} \citep{dimartino2014autism} and schizophrenia on
\textbf{COBRE} \citep{calhoun2012}; drug consumption on
\textbf{ACPI}; IQ measures on 
\textbf{HCP} \citep{vanessen2013hcp}; and age (with a regression model)
in normal aging with
\textbf{CamCAN} \citep{taylor_2017}.

\subsection{Quality of image reconstruction}\label{sec:image-rec-pipeline}

The signals extracted on a brain atlas can be seen as a compression, or
simplification, of the original signal. Indeed, a full image can be
reconstructed from these signals. We quantify
the signal loss incurred by this reduction.
For this, we project a brain map $\x$ onto an atlas (solving Eq. \eqref{eq:reduction}), and compute the best reconstruction of $\x$ from the loadings $\bal$, namely $\hat \x  = \D \bal \in \RR^p$.
We compare original and reconstructed images through the $R^2$
coefficient,
\begin{equation}
    \label{eq:r_squared}
    R^2(\x, \hat \x) = 1 - \frac{\Vert \x - \hat \x \Vert_2^2}{\Vert \x - \bar \x \Vert_2^2},
\end{equation}
where $\bar \x \in \RR$ is the spatial mean of map $\x$. The $R^2$ coefficient is averaged
across all images. Higher $R^2$ coefficients means that the
reduced signals (IDPs) explain more variance of the original images,
where $R^2=1$ corresponds to no signal loss. The larger the number of signals
used, the easier it is to explain variance; it is therefore interesting to
compare this measure across atlases with similar number of components.

\paragraph{Data}

We use \textbf{NeuroVault} \citep{gorgolewski2015}, the largest
public database of statistical maps.
To avoid circularity, we exclude maps derived from the studies used to
extract the DiFuMo atlases, along with maps that fail semi-automated quality
inspection (filtering out thresholded or non-statistical maps), resulting
in \textbf{15,542} maps.

\paragraph{Meta-analysis of contrasts maps}\label{sec:meta-analysis-pipeline}

Ideally, the extracted IDPs should allow to compute meta-analytical
summaries of brain activity maps. In this setting, a single
map, corresponding to a certain cognitive concept, is computed
from many $z$-maps across different studies, associated to 
conditions that involve this cognitive concept. We compare the summaries
obtained at voxel-level, i.e. averaging the maps $\{\x\}$, with the ones obtained
using reconstructed images, i.e. averaging the maps $\{\tilde \x\}$ used in
Eq. \eqref{eq:r_squared}. We use maps from our curated subset of
NeuroVault annotated with terms \textit{motor, language} and \textit{face recognition}.


\begin{figure}[t!]
    \begin{minipage}{1.\linewidth}
            \centerline{%
   \includegraphics[width=0.8\linewidth]{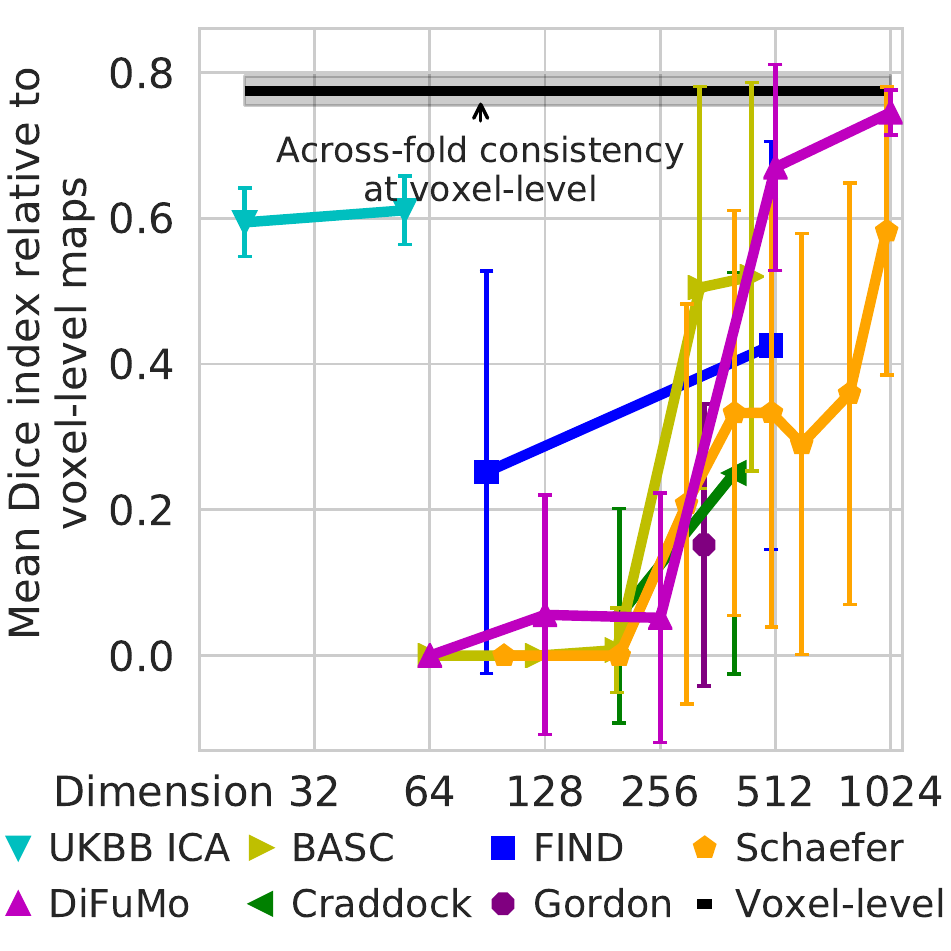}%
   \raisebox{46ex}{%
	   \llap{\rlap{\textbf{\sffamily General Linear Model on task fMRI}}
           \hspace{0.73\linewidth}}%
   }%
    }%
    \end{minipage}
    \begin{minipage}{1.\linewidth}
            \centerline{%
   \includegraphics[width=1.\linewidth]{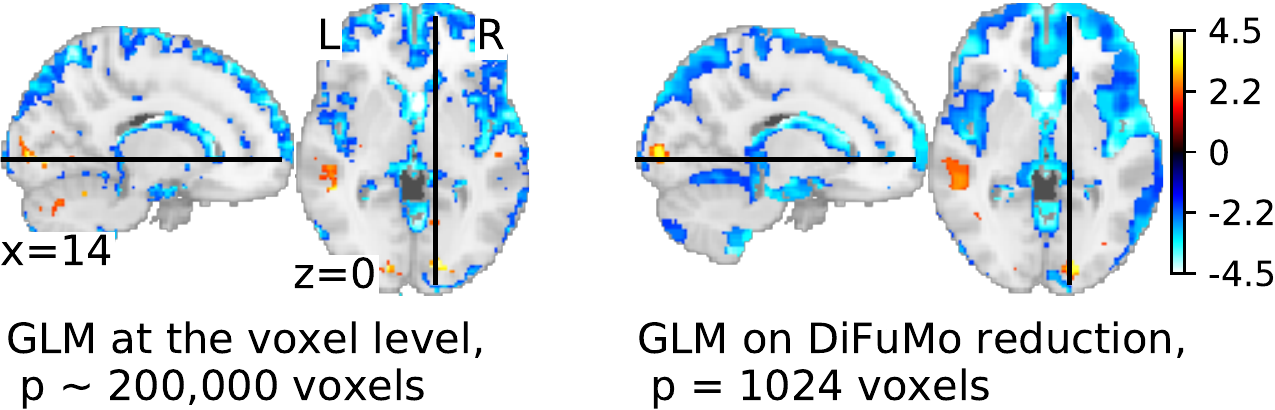}%
   \raisebox{19.3ex}{%
	   \llap{\rlap{\textbf{\sffamily Group-level $z$-map: complex sentence}}
           \hspace{0.78\linewidth}}%
   }%
    }%
    \end{minipage}
	\caption{\textbf{Overlap between GLM maps obtained with
        functional atlases and voxel-level analysis}.
	\textbf{Top:} The overlap is measured with the Dice
        similarity coefficient. The black line gives a baseline the mean
        overlap between voxel-level contrast maps
        over several random selections of sessions per subject.
        The figure gives Dice similarity scores between the GLM maps
        computed with signals extracted on functional atlases and at the
        voxel-level, after reconstruction of full $z$-maps and voxel-level
        thresholding with FDR control.
        The best similarity is achieved for highest dimensionality, though 1024-dimensional DiFuMo atlas largely dominates
        1000-dimensional Schaefer parcellation.
        Each point is the mean and the error bar denotes the
    standard deviation over contrast maps.
    \textbf{Bottom:} The activity maps
	encoded on 1024-dimensional space capture the same information and much
	smoother to voxel-level.}
        \label{fig:glm_dice}
\end{figure}

\section{Results: comparing atlases for analyses}\label{sec:results}

We report benchmarking results on the analytic tasks listed in the previous section.

\subsection{Brain mapping: standard task-fMRI analysis}

\autoref{fig:glm_dice} reports the results of standard analysis of task fMRI (GLM),
comparing analysis at the voxel-level with analyses on signals extracted
from functional atlases.
Best correspondence is obtained at highest dimensionality, as the regions
are finer. Notably, analysis with DiFuMo of dimensionality 1024 is markedly
closer to voxel-level analysis than using the largest alternative, 
the 1000-dimensional Schaefer parcellation. In addition,
the Dice index relative to the voxel-level gold standard is comparable to
the Dice index between runs of voxel-level GLM estimated across folds.
We note that using soft functional modes from only 55 ICA
components shows excellent results, comparable to those
obtained using the 1000 components Schaefer atlas. This stresses the 
benefit of \textit{continuous functional modes} for the
analysis of task responses.
Overall, standard task-fMRI analysis 
on signals derived from 512 or 1024-dimensional DiFuMo gives results
close to the voxel-level gold standard (\autoref{fig:glm_dice}
shows that the maps are also qualitatively similar).
\autoref{fig:encoding_r2} shows similar trends while comparing intra-subject
explained-variance maps, both qualitatively and quantitatively.
Dimension reduction have the additional benefit of
alleviating the burden of correcting for multiple comparisons.

\begin{figure}[t!]
    \centering%
   \includegraphics[width=0.8\linewidth]{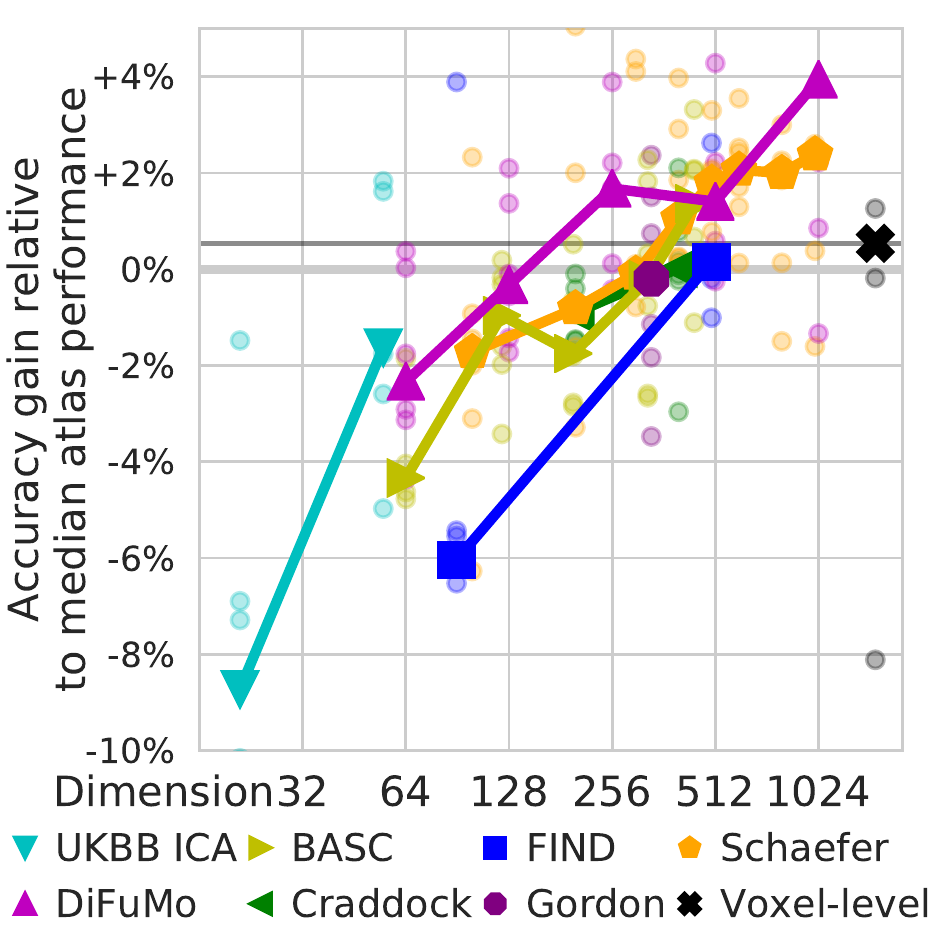}%
   \llap{\raisebox{.785\linewidth}{\textbf{\sffamily Decoding mental
processes from statistical maps}}}
    \caption{\textbf{Impact of the choice of atlas on decoding
	performance.}
	Each point gives the relative prediction score,
	over 6 different task-fMRI experiments.
	The thick lines give the median relative score
	per atlas.
    The baseline (black) is the relative score.
    High-order resolutions increase prediction accuracy.
	Using high-order DiFuMo ($k=1024$) and Schaefer parcellations
	($k=1000$) gives the best performance and, on average, outperforms
	voxel-level prediction.
    \ref{app:decoding_task_error_bars} reports absolute prediction scores
for each task separately.
}
    \label{fig:relative_median_task}
\end{figure}

\begin{figure}[t!]
{\bfseries\sffamily Voxel-level}\hfill%
{\bfseries\sffamily ~~~DiFuMo=1024}\hfill%
{\bfseries\sffamily Schaefer=1000}%

\includegraphics[trim={0 0.3cm 0 .75cm}, clip, width=.29\linewidth]{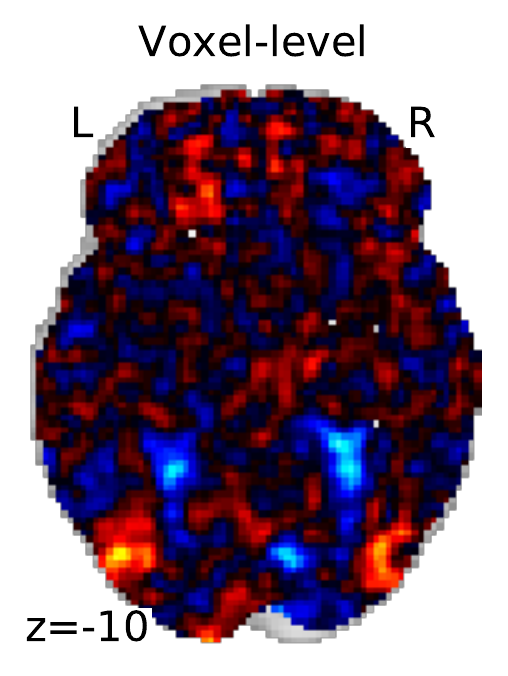}%
\hfill%
\includegraphics[trim={0 0.3cm 0 .75cm}, clip, width=.29\linewidth]{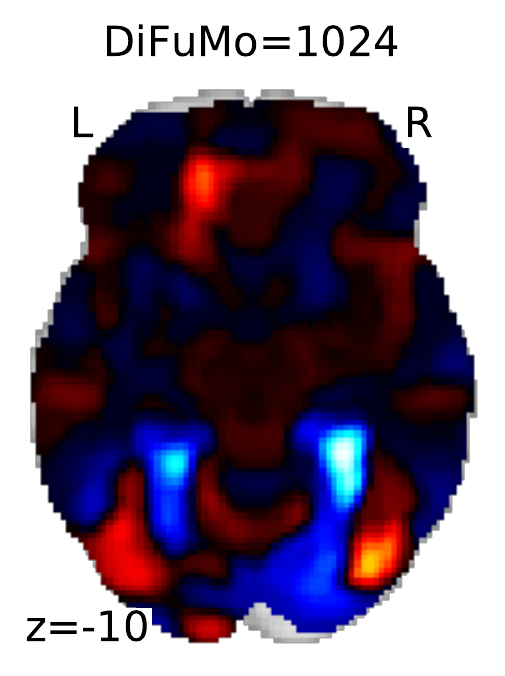}%
\hfill%
\includegraphics[trim={0 0.3cm 0 .75cm}, clip, width=.29\linewidth]{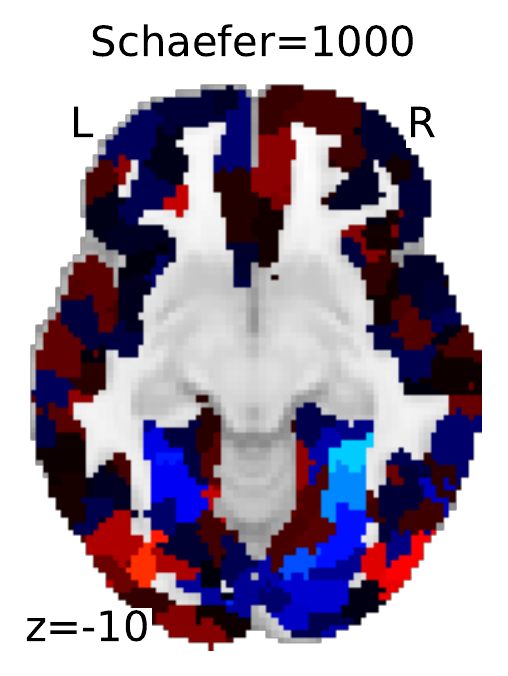}%

\caption{\textbf{Decoding maps of the working memory task, face
versus rest}, showed for Voxel-level analysis, DiFuMo, and Schaefer.
The maps are highly interpretable with
high-dimensional soft modes (DiFuMo 1024) compared
to voxel-level analysis. Brain areas important in the visual working
memory task --fusiform gyrus and
lateral occipital cortex-- are clearly visible. 
\autoref{fig:decoding_coefs} gives a full view of 
decoding weights across atlases and resolutions.
\label{fig:less_coefs}%
}
\end{figure}

\subsection{Decoding mental state from brain responses}

\autoref{fig:relative_median_task} shows the impact on decoding performance of 
reducing signals with various functional atlases.
It reports the performance relative to the median across methods for 
each of the 6 tasks.
These results clearly show the importance of high-dimensional functional modes
for decoding. Indeed, the higher the atlas resolution, the better the
predictions. Using DiFuMo $k=1024$ or Schaefer $k=1000$ gives the best
performance. In addition, as these functional atlases segment sufficiently-fine
regions, prediction from the corresponding signals tends to outperform
voxel-level prediction. Indeed, applying multivariate models to a larger number
of signals with a limited amount of data is more prone to overfitting---data
reduction acts here as a welcome regularization.
Qualitatively, brain maps containing decoding weights can be reconstructed.
With high-dimensional atlases, they are interpretable and
capture information similar to voxel-level analysis
(\autoref{fig:less_coefs}).

\begin{figure}[!t]
    \centering
   \includegraphics[width=0.8\linewidth]{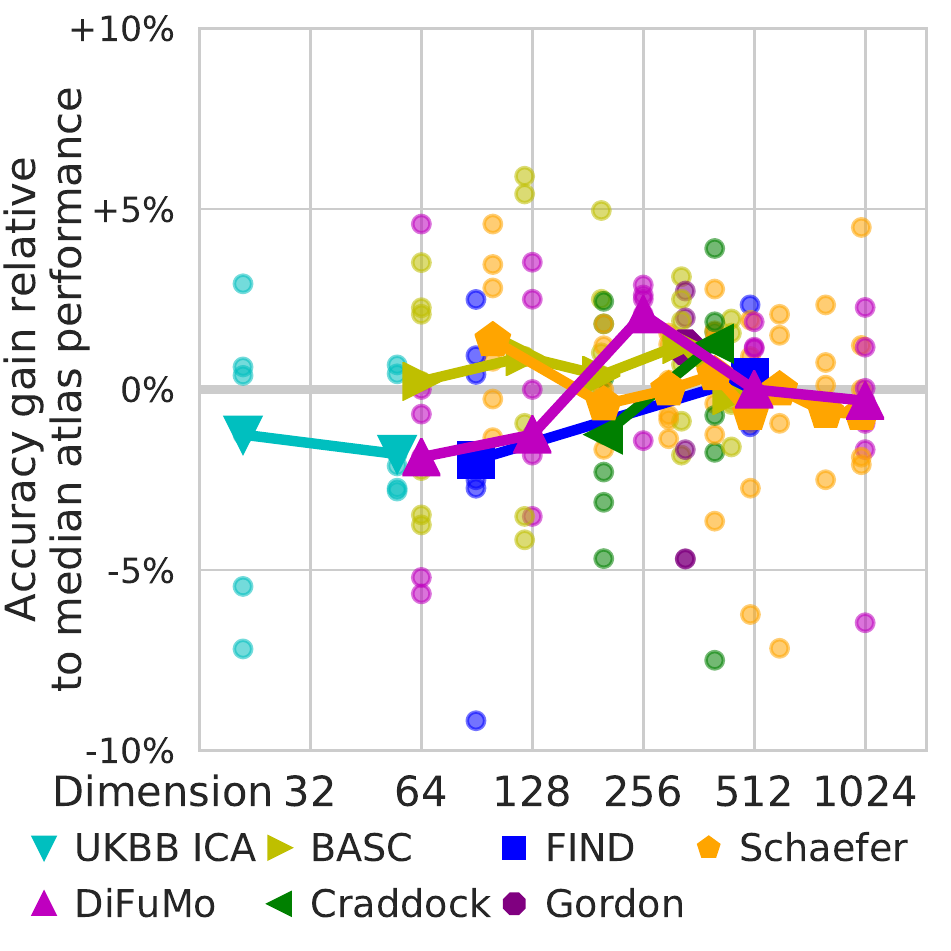}%
   \llap{\raisebox{.798\linewidth}{\textbf{\sffamily Predicting traits from functional connectomes}}}
    \caption[]{\textbf{Impact of the choice of atlas for predictions
	based on functional connectomes.}
	Each data point gives the prediction accuracy relative to the
	median for one of the 
	7 phenotypic prediction targets, i.e. each point a dataset.
	The thick line shows the median over the datasets.
	While the results are noisy, the optimal dimensionality seems to
	lie around 300 nodes, and the best-performing atlas is
	DiFuMo $k=256$, followed by Craddock $k=400$ and BASC $k=444$.
	\autoref{fig:decoding_rest_error_bars} report absolute results for each
	prediction problem.
}
    \label{fig:relative_median_rest}
\end{figure}

\subsection{Predicting traits from functional connectomes}

\begin{figure*}[t]
    \begin{minipage}{.3\linewidth}
    \caption{\textbf{Image reconstruction quality.} 
	    \textbf{Left:} Quantitative comparison on 15542 statistical
	    images. The $R^2$ loss between the
	    true and recovered images after compression with brain atlases of
	    multiple resolutions.
	    In general, higher-order atlases capture more signal.
	    \textbf{Right:} Meta-analysis summaries for the motor task.
	    High $R^2$ score (left) correspond to better capturing fine
	    structures of images, as visible on the qualitative images.
	    DiFuMo atlases better capture the gradients and smooth aspects
	    of the original images than hard parcellations, as BASC.
	    }
        \label{fig:compression}
    \end{minipage}
    \begin{minipage}{.3\linewidth}
	    \centerline{%
   \includegraphics[width=0.9\linewidth]{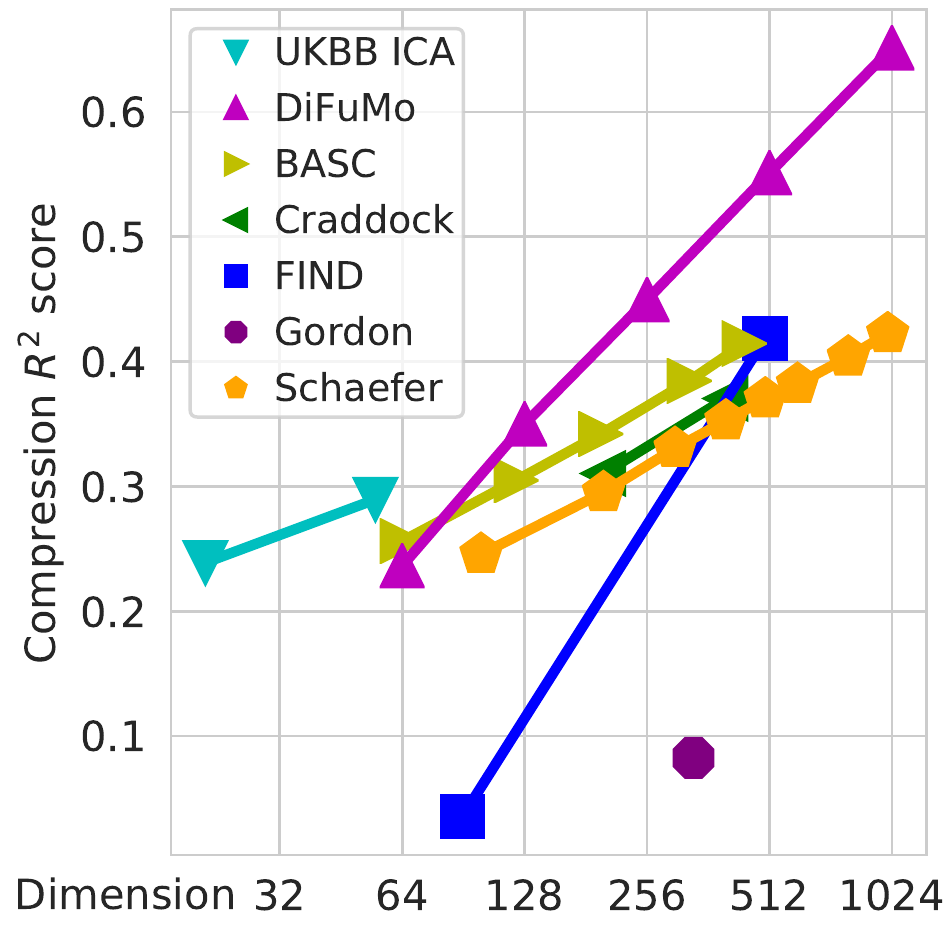}%
   \raisebox{35ex}{%
	   \llap{\rlap{\textbf{\sffamily Measuring data fidelity across many}}
	   \hspace{0.65\linewidth}}%
   }%
   \raisebox{33ex}{%
	   \llap{\rlap{\textbf{\sffamily statistical images}}
	   \hspace{0.1\linewidth}}%
   }%
    }%
    \end{minipage}
    \begin{minipage}{.5\linewidth}
	    \centerline{%
   \includegraphics[width=0.5\linewidth]{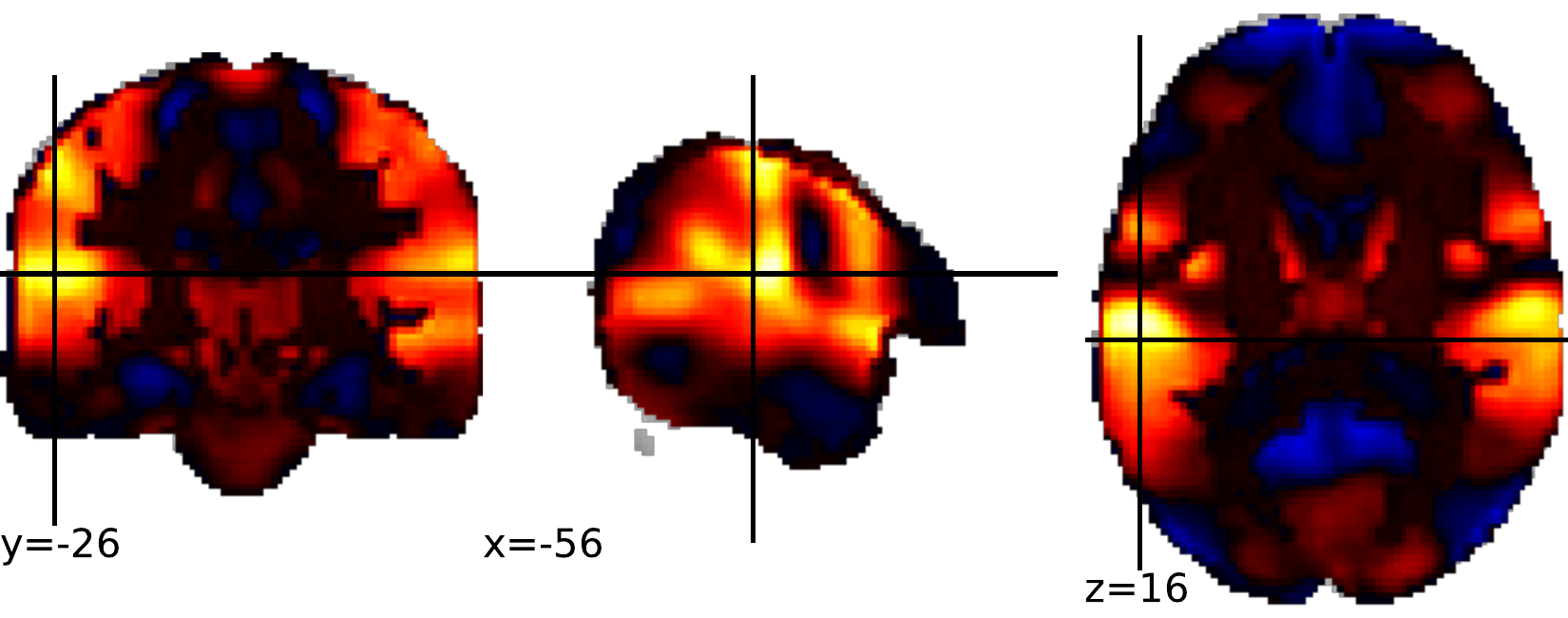}%
   \raisebox{-0.6ex}{%
       \llap{\rlap{\quad \sffamily Non-reduced image}
       \hspace*{0.76\linewidth}}%
   }%
    }%
	     \centerline{%
   \includegraphics[width=0.5\linewidth]{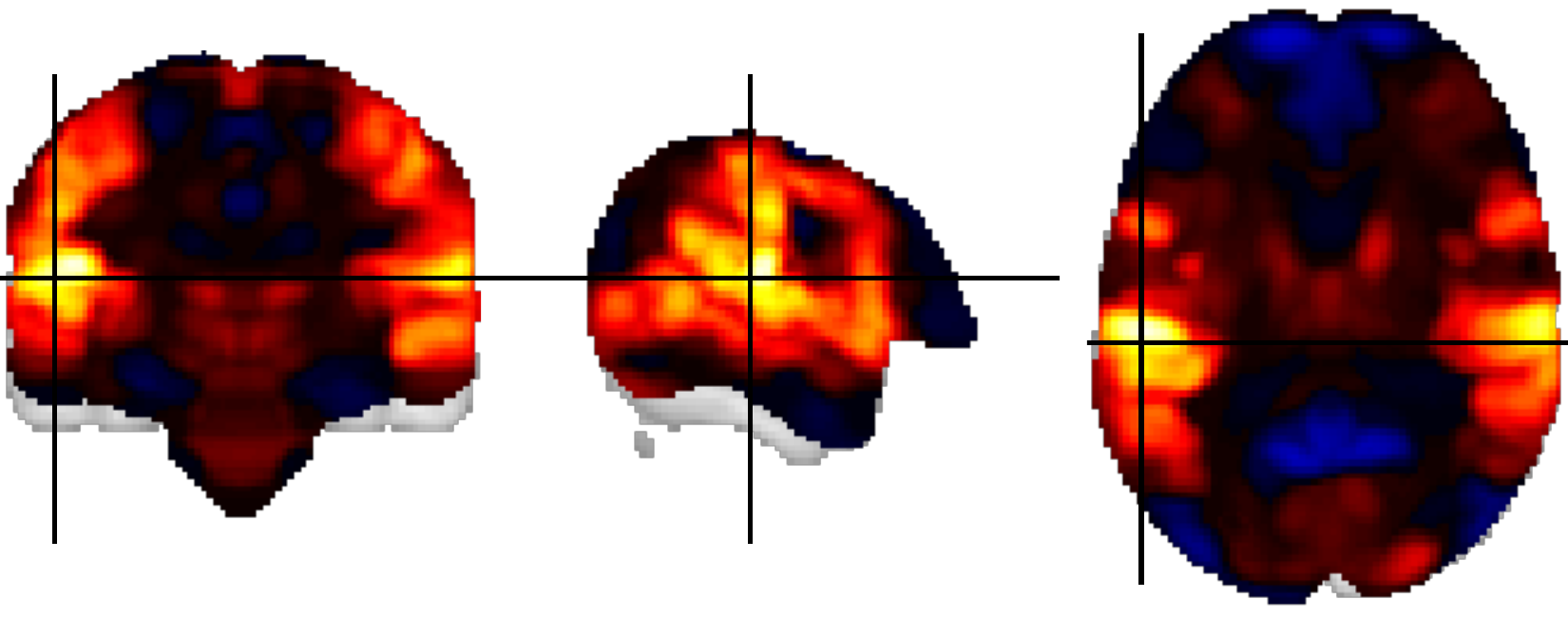}%
   \raisebox{0.01ex}{%
	   \llap{\rlap{\quad \sffamily Reduced with DiFuMo}
       \hspace*{0.76\linewidth}}%
   }%
    }%
	    \centerline{%
   \includegraphics[width=0.5\linewidth]{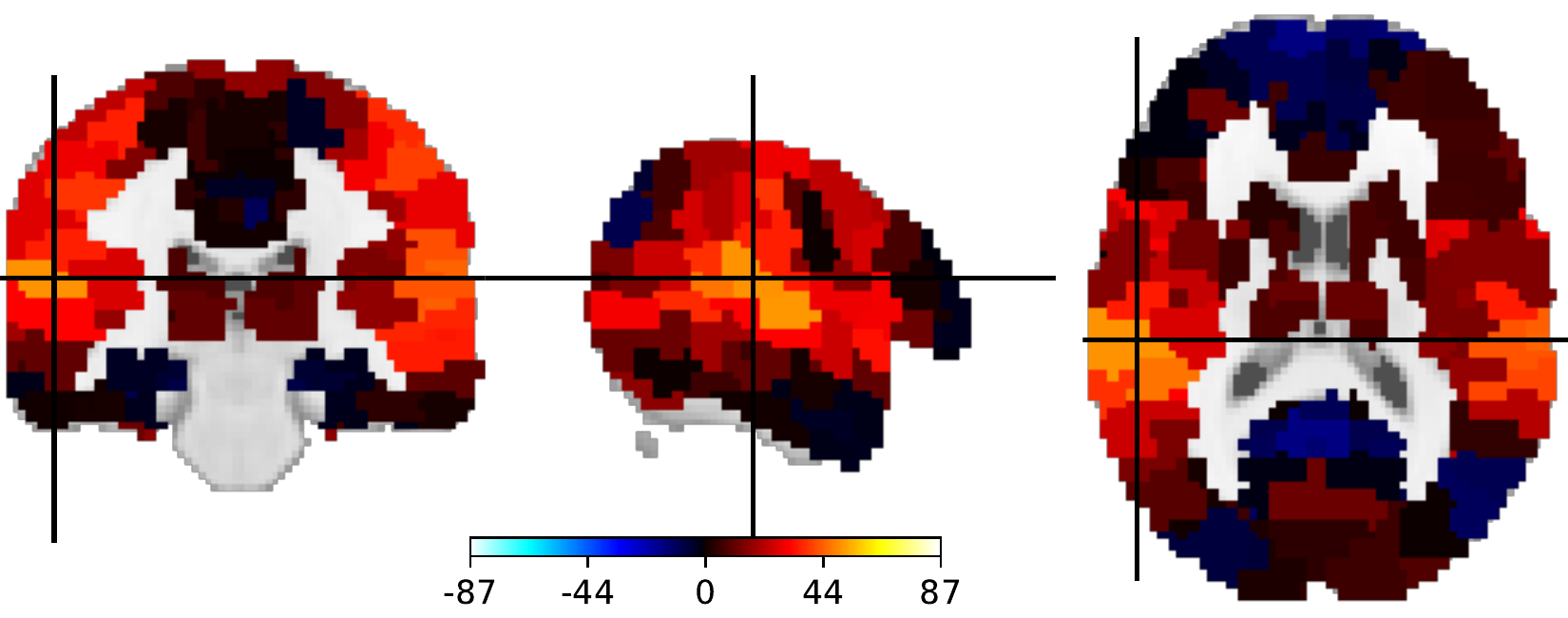}%
   \raisebox{0.01ex}{%
       \llap{\rlap{\quad  \sffamily Reduced with BASC}
       \hspace*{0.76\linewidth}}%
   }%
    }%
    \end{minipage}
\end{figure*}

\autoref{fig:relative_median_rest} shows the impact of the choice of
functional atlas when predicting phenotypes
from functional connectomes.
It reports the relative prediction accuracy for 7 different
prediction problems (each composed of a dataset and a target phenotype); the lines give the
median across the prediction problems.
Here, we do not report a voxel-level baseline, as it requires
to compute covariance matrices of dimensions around $100,000 \times
100,000$ and is therefore computationally and statistically intractable.
Unlike with the previous results, high-resolution atlases do not provide
the best performance, likely because the complexity of the statistical
models increases with the square of the number of nodes.
The best prediction overall is achieved using DiFuMo $k=256$, followed by
Craddock $k=400$ and BASC $k=444$ atlases.
Different outcomes 
have different optimal dimensionality, consistently across atlases
(\autoref{fig:decoding_rest_error_bars}): $k \sim 150$ for age
prediction; $k \sim 300$ for Autism Spectrum Disorder, PTSD,
or IQ prediction; and $k \sim 50$ 
for Alzheimer's Disease and drug use prediction.

\enlargethispage{1ex}%

\subsection{Fraction of the original signal captured}
\autoref{fig:compression} (left) displays the $R^2$ scores
summarizing the loss of information 
when data are reduced on an atlas and reconstructed back to
full images.
Unsurprising, reducing the images with lower-order dimensions
(atlases with fewer regions) yields a high loss of information
across all methods.
DiFuMo $k=1024$ captures $70\%$ of the original voxel-level signal.
Qualitatively, the
benefits of functional modes can be seen by comparing the meta-analytic maps
related to motor tasks (\autoref{fig:compression} right)---\autoref{fig:meta-analysis_more_cognitive_topics}
shows additional meta-analysis
on other topics. The DiFuMo have clear visual benefits over brain
discrete parcellations, such as BASC, as they better capture gradients.

%

\section{Discussion}\label{sec:discussion}

This paper introduces brain-wide soft functional
modes, named DiFuMos and made of a few hundreds to a thousand of brain
sub-divisions. They are derived from BOLD time-series across many
studies to capture well functional images with a small number of
signals.
In the context of population imaging, these signals are known as
image-derived phenotypes \citep[IDP,][]{miller2016} and are crucial to
easily scale statistical analysis, building a science of inter-individual
differences by relating brain signals to behavioral traits
\citep{dubois2016building}. Reducing the dimensionality of the signals
not only come with a $1000 \times$ gain in storage, but also with $100 \times$
computational speed-up for the analysis (\autoref{tab:computation_times}).
Even small-scale studies may need functional nodes,
e.g. for computing functional connectomes
\citep{zalesky2010_nodes,varoquaux2013connectomes}.
There already exist many functional brain atlases; yet DiFuMos have the unique advantage of being both soft and highly resolved. These features are important to capture gradients of functional
information.

\paragraph{Grounding better image-derived phenotypes} Signals extracted
from a functional atlas should
enable good statistical analysis of brain function. We
considered quantitative measures for typical neuroimaging analytic
scenarii and compared the fitness of extracting signal on DiFuMo with
using existing functional brain atlases. The biggest gains in analysis come
from increasing the dimensionality of brain sub-divisions, aside for
functional connectome studies where an optimal is found around 200 nodes.
Choosing the number of nodes then becomes a tradeoff between complexity
of the representation and analytic performance. Importantly, the gains in
analytic performance continue way beyond the dimensionality typically
used for IDPs \citep[e.g. $55$ components from][]{miller2016}.
These results extend prior literature emphasizing the importance of 
high-dimensional parcellations for fMRI
\citep{abou-elseoud2011, thirion2014, arslan2017, salallonch2019}.
To foster good analysis, the second most important aspect of a
parcellation appears that it be soft, i.e. continuously-valued. For a given
dimensionality, soft modes tend to outperform hard
parcellations, whether they are derived with ICA or dictionary learning.

\begin{figure*}[t!]
\begin{minipage}[b]{.285\linewidth}%
\caption{\textbf{Modes around the putamen},
for DiFuMo dimensionality 64, 256, and 512.
As dimensionality increases: sub-divisions are more refined, modes
are split into right and left hemisphere and
anterio-posterior direction. Each color represents a single mode.
\autoref{fig:putamen_supp} details more
this breakdown.\label{fig:putamen}%
}%
\end{minipage}%
\hfill%
\begin{minipage}[b]{.23\linewidth}
\includegraphics[trim={.5cm .5cm 1.05cm .45cm}, clip, width=\linewidth]{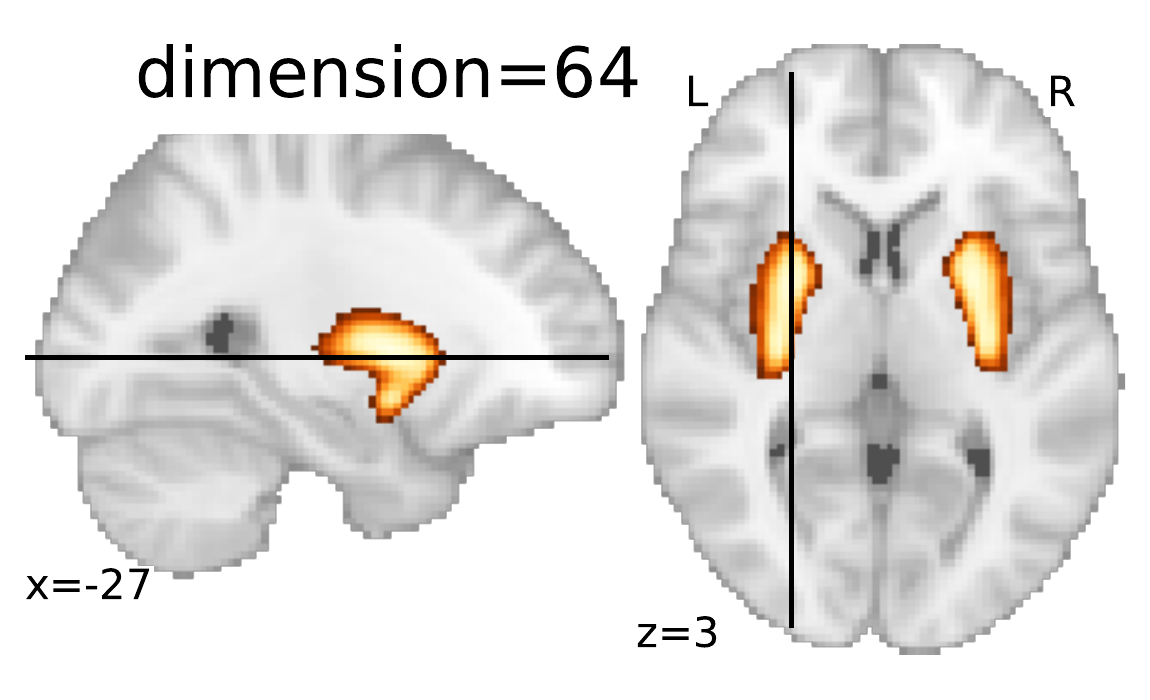}%
\llap{\raisebox{1.2ex}{\rlap{\sffamily\small 1 mode}\hspace*{.75\linewidth}}}%
\end{minipage}%
\hfill%
\begin{minipage}[b]{.23\linewidth}
\includegraphics[trim={.5cm .5cm 1.05cm .45cm}, clip, width=\linewidth]{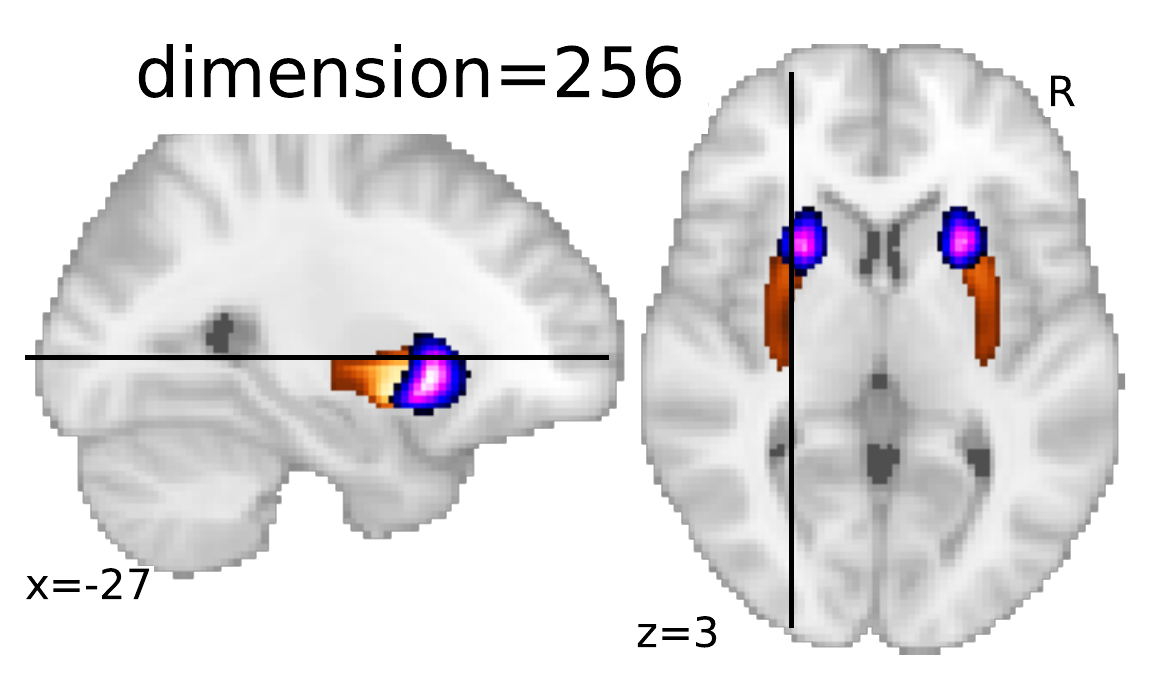}%
\llap{\raisebox{1.2ex}{\rlap{\sffamily\small 2 modes}\hspace*{.75\linewidth}}}%
\end{minipage}%
\hfill%
\begin{minipage}[b]{.23\linewidth}
\includegraphics[trim={.5cm .5cm 1.05cm .45cm}, clip, width=\linewidth]{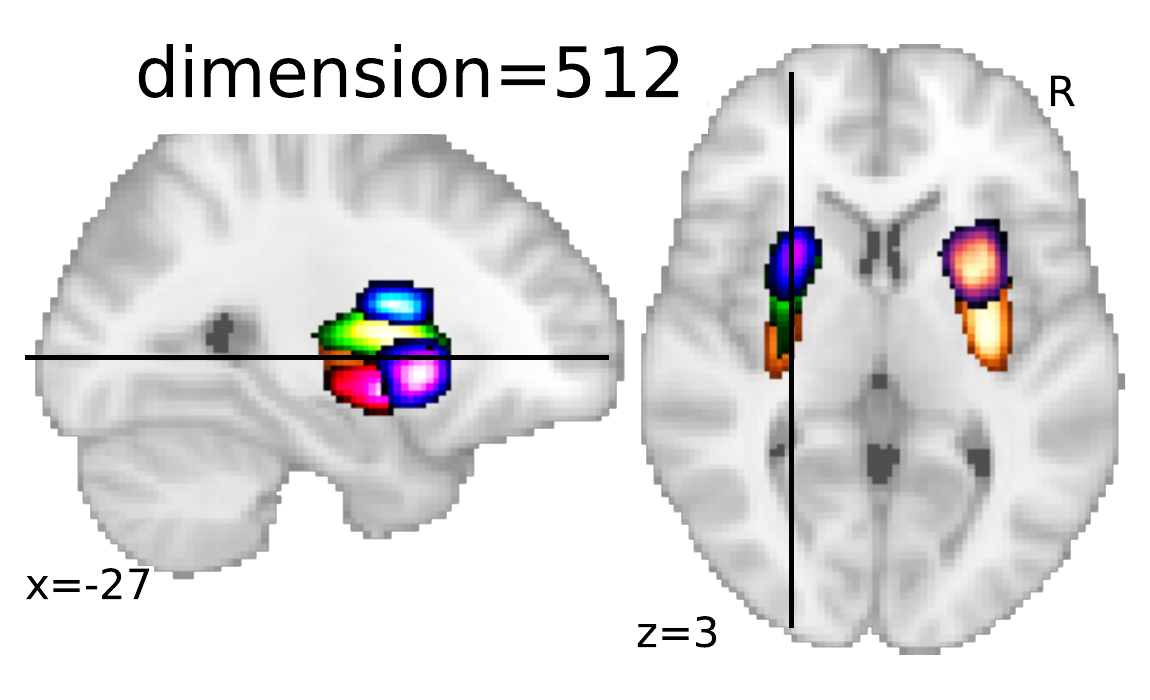}%
\llap{\raisebox{1.2ex}{\rlap{\sffamily\small 6 modes}\hspace*{.75\linewidth}}}%
\end{minipage}%
\end{figure*}

\paragraph{Modes well-adapted to the EPI signal}
The functional modes are optimized to fit well a large number of EPI
images: 2,192 sessions across 27 studies. As a result, they form a division of the brain
well adapted to the signal. For instance, they define regions 
larger in the white matter and in the CSF than in the grey
matter (\autoref{fig:region_sizes_in_cm3}).
A large dataset is needed to capture such implicit regularities of
the signal with high-dimensional spatial decompositions. Indeed, running
the same model on less data extracts modes with less spatial regularity
(\autoref{fig:comparisons_on_training_sizes}).
The combination of high dimensionality and large dataset leads
to significant computational demands. The extraction of DiFuMos was
possible thanks to fast algorithms for huge matrix factorization
\citep{mensch2018ieee}, and gathering
data representative of a wide variety of scanning
protocols via openfMRI \citep{poldrack2013openfmri}.

We did not limit the DiFuMo modes to gray matter, as
measures outside gray matter can be useful in subsequent analysis, for
instance to remove the global signal \citep{murphy2017towards}. In
addition, distributed modes extracted from full-brain EPI can separate
out noise --such as movement artifacts-- and help rejecting it in a
later analysis \citep{perlbarg2007,griffanti2014ica,pruim2015evaluation}.
Some DiFuMo modes indeed segment ventricles or 
interfaces. Depending on the application, practitioners can choose
to restrict signal extraction to a grey-matter mask.

\paragraph{The functional modes are sharp and anatomically relevant}

To extract structures defined by brain anatomy or microstructure,
atlasing efforts have used anatomical or multimodal imaging
\citep{mori2005mri,desikan2006,eickhoff2007assignment,glasser2016}.
The DiFuMo atlases capture a different signal: brain activity.
Yet, thanks to the sparsity and non-negativity constraint, they are made of localized modes
which often have a natural anatomical interpretation.
Consequently, we have labeled the modes with a unique name based on the
most relevant anatomical structure, following \cite{urchs2019mist} who
also give anatomical labels to functional regions.
Indeed, using a common vocabulary of brain structures is important for
communication across the neuroimaging community.
As visible on \autoref{fig:putamen}, the modes are well anchored on
anatomical structures such as the putamen.
They are however not constrained to contain only one connected region.
Smaller dimension DiFuMos indeed capture distributed networks, often comprising
bilateral regions. As the dimensionality increases, the networks
progressively separate in smaller networks which eventually form single
regions. For instance, the left and right putamen appear in the same mode 
at dimension 64, and are first sub-divided along the anterio-posterior
direction, and later the left and right putamen are separated
(\autoref{fig:putamen}). Dimension choice is data driven: it should best explain the functional signal.



\section{Conclusion}

We provide multidimensional atlases of functional modes that can be
used to extract functional signals:
\href{https://parietal-inria.github.io/DiFuMo}{parietal-inria.github.io/DiFuMo}.
They give excellent performance for a wide variety of
analytic tasks: GLM-based analysis, mental-process decoding or
functional-connectivity analysis.
Their availability reduces computational burdens: practitioners
can readily perform analyses on a reduced signal, without a
costly ROI-definition step.
In addition, working on common functional modes across studies
facilitates comparison and interpretations of results.
To help communication, we have
labeled every functional mode to reflect the
neuroanatomical structures that it contains.
To date, these are the only high-dimensional soft
functional modes available.
As they have been extracted from a variety of
data (more than 2,000 sessions across 27 studies, 2.4TB in size)
and improve many
analytic tasks, the rich descriptions of neural activity that they
capture is
well suited for a broad set of fMRI studies.

\section{Acknowledgments}
This project has received funding from the European Union’s Horizon 2020
Research and Innovation Programme under Grant Agreement No. 785907 (HBP SGA2)
and No 826421 (VirtualBrainCloud). The work of Arthur Mensch has been
supported by the European Research Council (ERC Grant Noria). This work acknowledges the support of ANR NeuroRef and ERC-StG NeuroLang.

We also thank Pierre Bellec and Vincent Frouin for their helpful
discussions on the experimental work, the neuroimaging
community for giving access to fMRI datasets, and open-source
contributors on the packages we build upon (including \textit{nilearn}, \textit{fMRIprep},
and \textit{MRIQC}).

Data collection and sharing for this project was funded by the Alzheimer's Disease Neuroimaging Initiative
(ADNI) (National Institutes of Health Grant U01 AG024904) and DOD ADNI (Department of Defense award
number W81XWH-12-2-0012). ADNI is funded by the National Institute on Aging, the National Institute of
Biomedical Imaging and Bioengineering, and through generous contributions from the following: AbbVie,
Alzheimer's Association; Alzheimer's Drug Discovery Foundation; Araclon Biotech; BioClinica, Inc.; Biogen;
Bristol-Myers Squibb Company; CereSpir, Inc.; Cogstate; Eisai Inc.; Elan Pharmaceuticals, Inc.; Eli Lilly and
Company; EuroImmun; F. Hoffmann-La Roche Ltd and its affiliated company Genentech, Inc.; Fujirebio; GE
Healthcare; IXICO Ltd.; Janssen Alzheimer Immunotherapy Research and Development, LLC.; Johnson and Johnson Pharmaceutical Research and Development LLC.; Lumosity; Lundbeck; Merck and Co., Inc.; Meso
Scale Diagnostics, LLC.; NeuroRx Research; Neurotrack Technologies; Novartis Pharmaceuticals
Corporation; Pfizer Inc.; Piramal Imaging; Servier; Takeda Pharmaceutical Company; and Transition
Therapeutics. The Canadian Institutes of Health Research is providing funds to support ADNI clinical sites
in Canada. Private sector contributions are facilitated by the Foundation for the National Institutes of Health
(www.fnih.org). The grantee organization is the Northern California Institute for Research and Education,
and the study is coordinated by the Alzheimer's Therapeutic Research Institute at the University of Southern
California. ADNI data are disseminated by the Laboratory for Neuro Imaging at the University of Southern
California.



\let\oldbibitem\bibitem
\def\bibitem{\vfill\smallskip\nopagebreak\oldbibitem}

\bibliography{biblio}

\appendix
\renewcommand{\thefigure}{A\arabic{figure}}
\setcounter{figure}{0}
\renewcommand{\thetable}{A\arabic{table}}
\setcounter{table}{0}


\begin{figure}[t!]
   \hspace*{.05\linewidth}%
   \begin{minipage}{0.5\paperwidth}
     \includegraphics[width=0.8\linewidth]{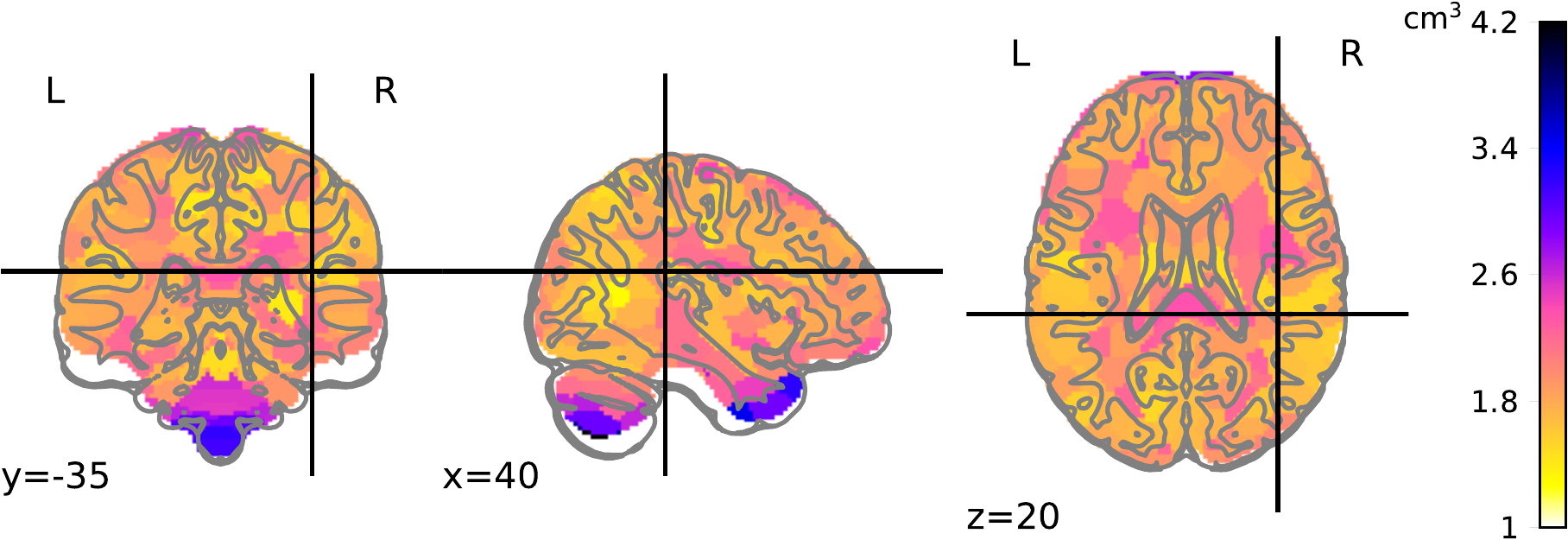}%
     \llap{\raisebox{.28\linewidth}{%
   \parbox{0.82\linewidth}{\bfseries\sffamily DiFuMo atlases capture well the
	                                      EPI signal}}}%
    \end{minipage}
	\caption{\textbf{Region volume ($cm^{3})$ of modes on the brain with 1024
	dictionary of DiFuMo.} The volume of the modes tends to be larger
	corresponding to white matter when compared with the cortical gray
	matter. This justifies the adaptation of DiFuMo atlas to the fMRI
	signal.}
   \label{fig:region_sizes_in_cm3}
\end{figure}

\begin{figure}[tt]
   \begin{minipage}{.66\linewidth}
	\begin{tikzpicture}[spy using outlines={green, circle,
       magnification=1.8, size=70, connect spies}]
     \node{\includegraphics[width=.7\linewidth]
       {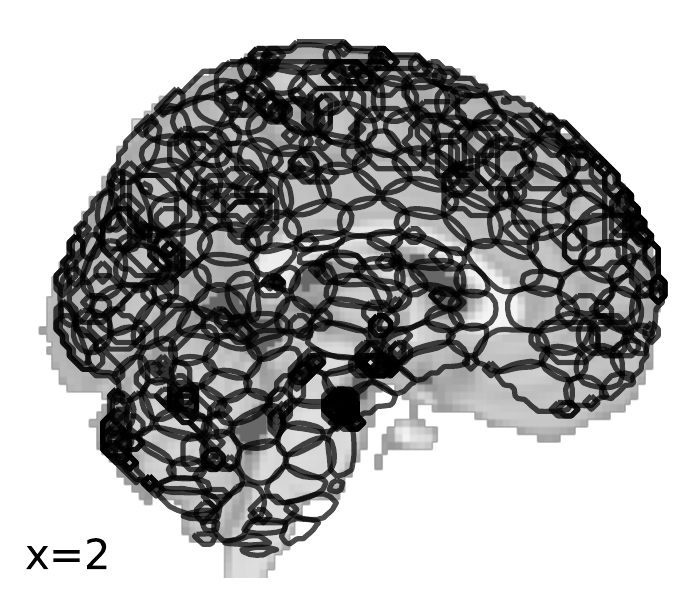}};
     \spy on (-0.5, 0.55) in node[overlay, line width=2mm]
       at (4.1, 0.2);
     \end{tikzpicture}%
       \llap{\rlap{\raisebox{.62\linewidth}{%
           \bfseries\sffamily $1/20$th of full training size
       }}\hspace*{.55\linewidth}}%
   \end{minipage}
   \begin{minipage}{.66\linewidth}
	\begin{tikzpicture}[spy using outlines={green, circle,
       magnification=1.8, size=70, connect spies}]
     \node{\includegraphics[width=.7\linewidth]
       {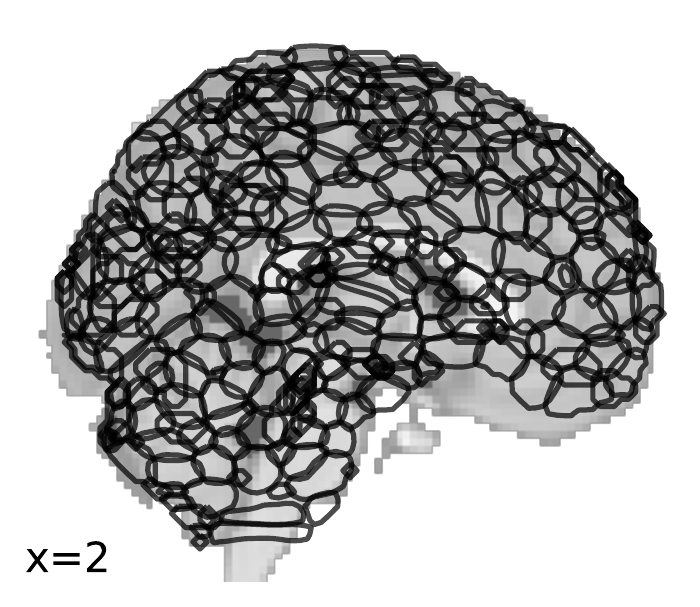}};
     \spy on (-0.5, 0.55) in node[overlay, line width=2mm]
	at (4.1, 0.2);
     \end{tikzpicture}%
       \llap{\rlap{\raisebox{.62\linewidth}{%
           \bfseries\sffamily Full training size: 2192 volumes
       }}\hspace*{.55\linewidth}}%
   \end{minipage}
	\caption{1024 components trained on two different sizes of the input set of fMRI images. The components trained on the full data have
	    more spatial regularity, while the components trained on 100
volumes have more overlap in some regions of the brain. The
additional spatial regularity shows the benefit of large-scale training size
	   in learning a data-driven based functional atlas.}
	   \label{fig:comparisons_on_training_sizes}
\end{figure}

\begin{table}[t]
    \small
    \begin{center}
        \begin{tabular}{l|l|l|l|l}
            \hspace*{-0.1em}\rotatebox{30}{\bf Task} &
\hspace*{-0.45em}\rlap{\raisebox{3ex}{\rotatebox{30}{\bf
\#}}}\rotatebox{30}{\bf samples}\hspace*{-0.45em} &
            \hspace*{-0.5em}\rotatebox{30}{\bf Representation} &
\hspace*{-0.5em}\raisebox{2ex}{\rotatebox{30}{\bf
Time}}\hspace*{-1.6em}\rotatebox{30}{(sec)}\hspace*{-.5em}&
            \hspace*{-0.5em}\rotatebox{30}{\bf Speedup}
\\
            \hline\\[-4mm]
            \multirow{2}{*} {Emotion} & \multirow{2}{*}{4924} &
Voxel-level  & 77.7 &\multirow{2}{*}{$46\times$}\\
                                      & &\cellcolor{gray!25} Reduced
&\cellcolor{gray!25} 1.7 &\\
            \hline\\[-4mm]
            \multirow{2}{*} {Pain} & \multirow{2}{*}{84} & Voxel-level  &
1.5&\multirow{2}{*}{$250\times$}\\
                                      & &\cellcolor{gray!25} Reduced
&\cellcolor{gray!25} 0.006 &\\
            \hline\\[-4mm]
            \multirow{2}{*}{Working} & \multirow{2}{*}{3140} &
Voxel-level  & 874.7 &\multirow{2}{*}{$240\times$}\\
            {memory}  & &\cellcolor{gray!25} Reduced &\cellcolor{gray!25}
3.7 &\\
            \hline\\[-4mm]
            \multirow{2}{*} {Gambling} & \multirow{2}{*}{1574} &
Voxel-level & 298.7&\multirow{2}{*}{$270\times$}\\
                                      & &\cellcolor{gray!25} Reduced
&\cellcolor{gray!25} 1.12&\\
            \hline\\[-4mm]
            \multirow{2}{*} {Relational} & \multirow{2}{*}{1572} &
Voxel-level & 263.1&\multirow{2}{*}{$405\times$}\\
                                      & &\cellcolor{gray!25} Reduced
&\cellcolor{gray!25} 0.65&\\
        \end{tabular}
    \end{center}
    \caption{The comparison in computational times while predicting mental state
        on two set of brain features space: voxel-level $\approx 200,000$ and
        reduced voxels to DiFuMo 1024. We report the averaged time over 20
        cross-validation folds for several task-fMRI conditions.
        Clearly, there are benefits trading for reduced representations in terms
        of computation time. On high-resolution brain images like HCP, these
        are decreased by a factor $200$.}
        \label{tab:computation_times}
\end{table}

\section{Performance of DiFuMos}

As discussed in \autoref{sec:discussion}, we report how DiFuMOs components are
well adapted to the fMRI EPI signal in \autoref{fig:region_sizes_in_cm3}.
\autoref{fig:comparisons_on_training_sizes} qualitatively compare components
obtained training on the whole data corpus and training on a fraction of it.
Better component regularity is obtained with more data. Finally,
\autoref{tab:computation_times} reports the computational speed-ups obtained
using DiFuMos IDPs instead of voxel in the decoding experiment. Similar
speed-ups are observed in the other validation pipelines.

\section{Details on stimulus decoding}

We provide additional details for the decoding pipeline, to
complete the description in \autoref{sec:decoding-pipeline}.

\begin{table}[t!]
    \small
    \begin{center}
        \begin{tabular}{c|c|c}
            \hspace*{0.3em}\cellcolor{gray!25}\bf{Task-fMRI} &
            \hspace*{0.2em}\cellcolor{gray!25}\bf{Prediction task} &
            \hspace*{0.5em}\cellcolor{gray!25}\bf{\# maps} \\
            \hline\\[-3.8mm]
            NV503: Emotion & Rating:$1, 2, 3, 4, 5$ & 4924 \\
            \hline\\[-3.8mm]
            NV504: Pain    & Sensitivity: $1, 2, 3$       & 84 \\
            \hline\\[-3.8mm]
            HCP: Working mem. & face vs place   & 3140 \\
            \hline\\[-3.8mm]
            HCP: Gambling       & loss vs reward  & 1574 \\
            \hline\\[-3.8mm]
            HCP: Relational     & relational vs matching & 1572 \\
            \hline\\[-3.8mm]
            IBC: Archi standard & left vs right hand & 1040 \\
        \end{tabular}
    \end{center}
    \caption{\textbf{Dataset, prediction tasks and dataset size for each of the 6 decoding tasks we consider           in \autoref{sec:decoding-pipeline}.}
         $z$-maps from HCP and IBC were computed using the GLM, while NeuroVault directly provided the $\beta$-maps for Emotion and Pain. NV: NeuroVault.
        \label{tab:task-data-validation}}
\end{table}

\begin{figure}[t!]
    \centerline{%
        \includegraphics[width=1.\linewidth]{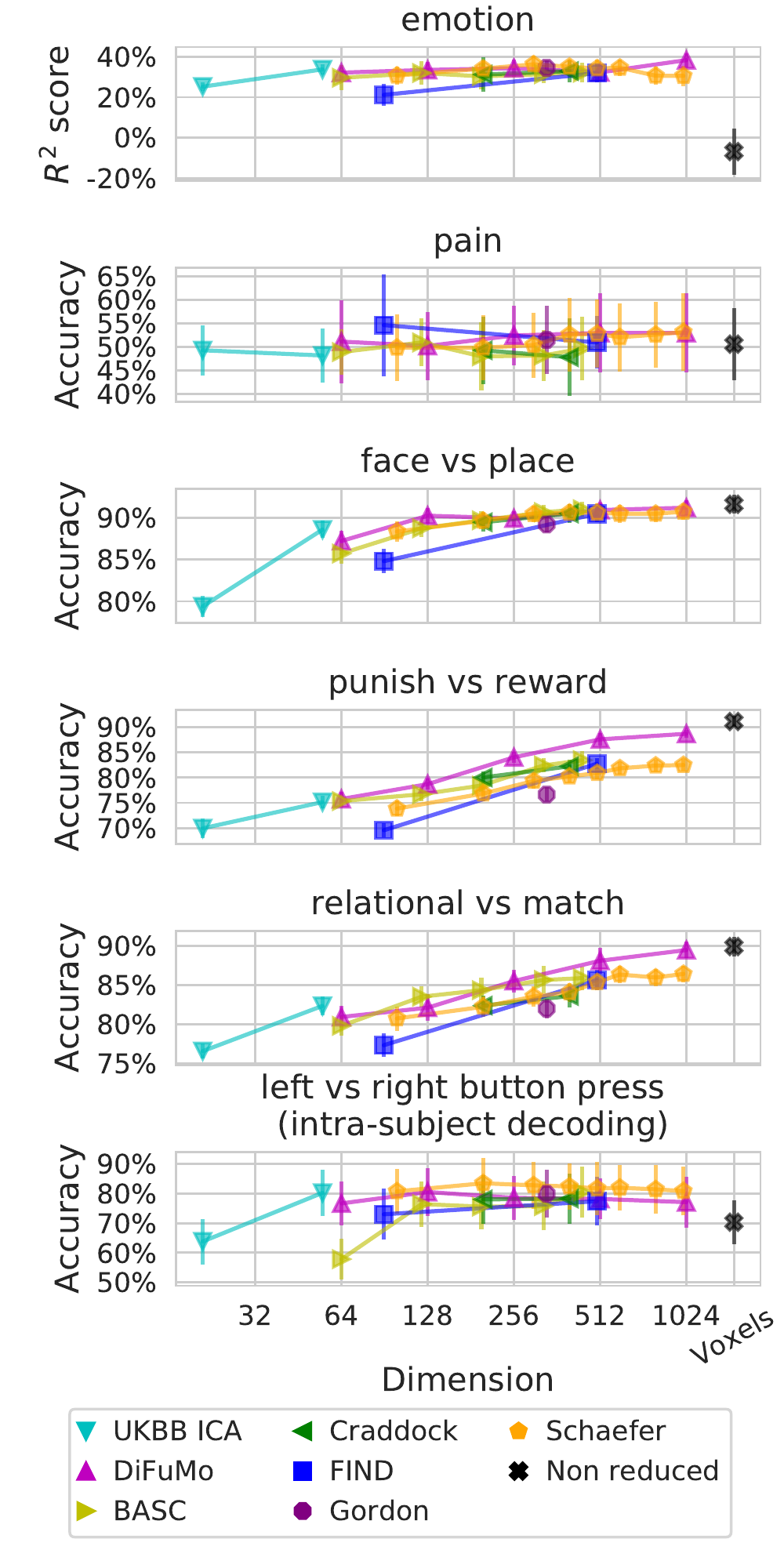}%
   \raisebox{118ex}{%
       \llap{\rlap{\textbf{\sffamily Predicting mental state:
       Task-fMRI }}\hspace*{0.9\linewidth}}%
   }%
    }%
    \caption{\textbf{Decoding prediction scores for each brain atlas and target}:
    Each marker denotes the mean performance of using a certain brain atlas; error bars are the standard deviation
    of the prediction scores for this atlas. Decoding from high-order dictionaries, and especially from DiFuMos, perform similarly or better than decoding from voxels.
    \label{fig:decode_task_error_bars}
    }
\end{figure}

\subsection{Input data and pre-processing pipelines}
\label{sec:task_fMRI_data}
The decoding pipeline classifies input unthresholded statistical maps.
\autoref{tab:task-data-validation} summarizes the task-based
studies used to obtain these statistical maps.

\paragraph{Pre-encoded maps downloaded from Neurovault.org}
We download maps related to emotion and pain \citep{chang2015emotion} using
\textit{Neurovault}, querying the collections \textit{503} and \textit{504}. We use the ``Rating'' \& ``PainLevel'' labels as predictive
targets. We predict emotion using ridge regression, and pain-level over 3 classes
using Linear SVC. The supervised learning pipeline, that includes
cross-validation and linear models is implemented with Python based
\textit{scikit-learn} \citep{pedregosa2011}. We use \textit{nilearn} \citep{abraham2014} to
download maps from Neurovault.org interface \citep{gorgolewski2015}. The data
acquisition parameters, preprocessing details and estimation of statistical
maps are described in \cite{chang2015emotion}.

\paragraph{Statistical maps encoded using the GLM} We compute $z$-maps from
HCP900 \citep{vanessen2012hcpdata} and IBC \citep{pinho2018ibc} studies,
that comprise high-qualiy task-fMRI experiments.

\textit{HCP.} We download fMRI data from the HCP900 release; those are already
preprocessed using HCP pipelines \citep{glasser2013}. We use MNINonLinear-based
registered data as input for the GLM, that outputs one $z$-map per condition per
subject. We consider three task-based studies, namely: for \textbf{Working
Memory}, we consider $z$-maps based on condition: ``0-back faces'', ``2-back
faces'', ``0-back places'', ``2-back places''. Similarly, for \textbf{Gambling}
\citep{delgado2000gambling}, we consider $z$-maps for the conditions ``loss''
and ``reward''; finally, on \textbf{Relational processing}, we consider $z$-maps
for the conditions ``relational processing'' and ``matching''. For each study,
we use Linear SVC on encoded $z$-maps to predict psychological conditions. The predictive
model therefore perform a 2-class or 4-class classification. The experimental
protocol and data acquisition parameters are detailed in
\cite{vanessen2012hcpdata}.

\textit{IBC.} We consider the \textbf{Archi Standard}
\citep{pinel2007} motor task, where subjects are asked to press ``left'' and
``right'' button press based on audio and visual instructions. We perform
within-subject classification between left and right button press, using
$z$-maps corresponding to each repetition of the instruction. For each of the 13 available
subjects, a linear model is trained on the $z$-maps from all but one session and
prediction is performed on the left-out session. Each subject provides 80
encoded $z$-maps across 4 sessions. We use data preprocessed following the
pipelines of \cite{pinho2018ibc}.

\textit{GLM specification.} For both datasets, the input $z$-maps are estimated from the raw fMRI data by fitting a GLM. We use Nistats\footnote{\url{https://nistats.github.io/}},
a Python package for the statistical analysis of fMRI data. The temporal
regressors of the model are specified according to the timing of stimulus
presentations convolved with hemodynamic models (\textit{spm + derivative}).
We use polynomial model to capture the low-frequency drifts in the data.

\begin{figure}[t!]
    \centerline{%
        \includegraphics[width=1.\linewidth]{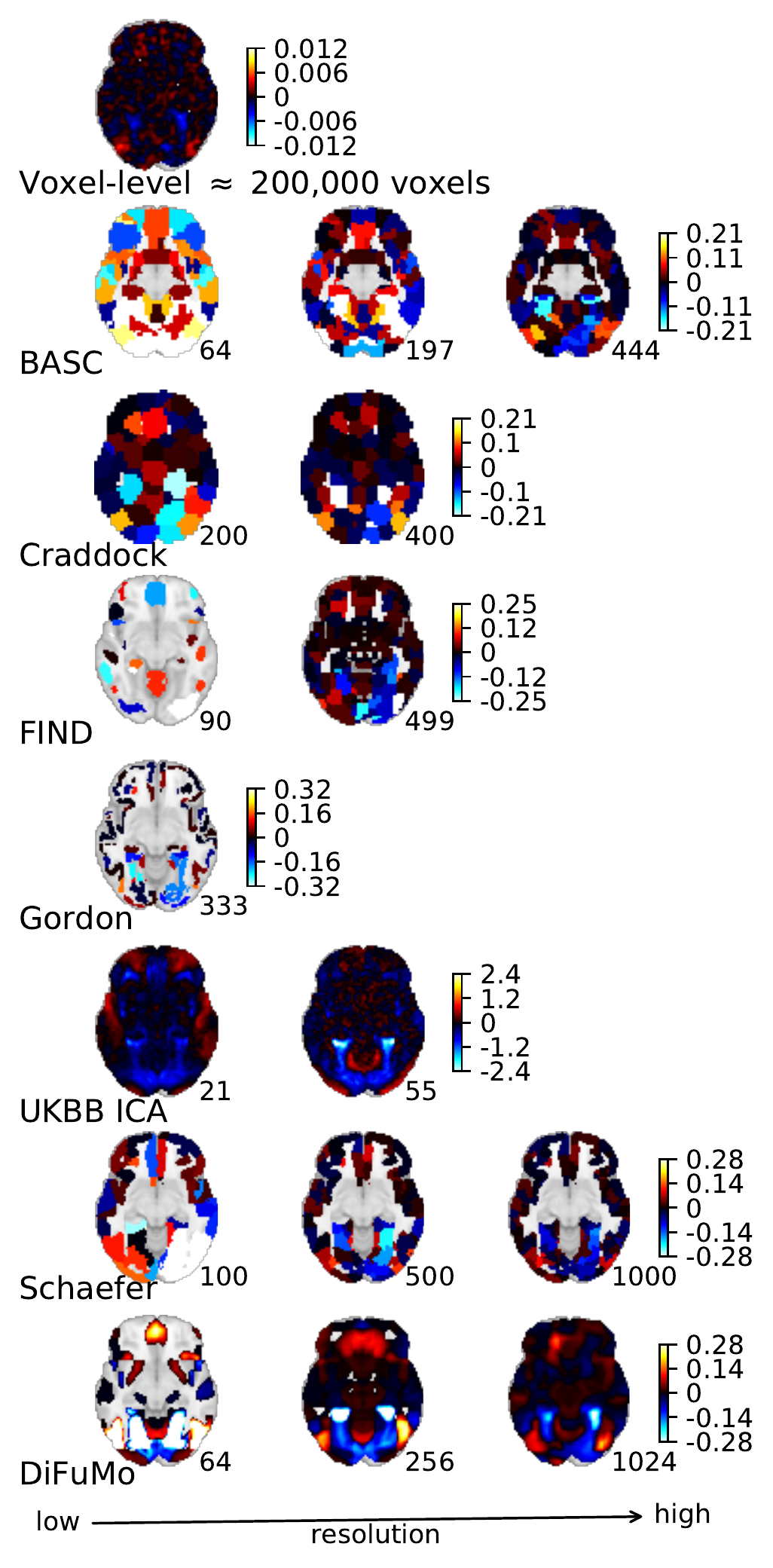}%
       \raisebox{120ex}{%
           \llap{\rlap{\textbf{\sffamily Decoding working memory: 0BK face versus rest}}\hspace*{0.93\linewidth}}%
       }%
    }%
        \caption{\textbf{Decoding classification weight maps for the HCP working memory task
        (0BK face)}, obtained with
        voxel-level decoding and decoding over various functional atlases. Using DiFuMos
        yield highly interpretable weight maps; it clearly delineates the fusiform
        gyrus and lateral occipital cortex.
    \label{fig:decoding_coefs}
    }
\end{figure}

\subsection{Detailed results}
\label{app:decoding_task_error_bars}

To complete the summarizing \autoref{fig:relative_median_task}, we report the
raw prediction scores separately for each decoding tasks in
\autoref{fig:decode_task_error_bars}. 
Prediction accuracy increases with the size of functional atlases. Using 1024
atlases allows to match or pass the performance of voxel-based prediction.
In terms of interpretation, the weights are much smoother and blobs are clearly
visible in the weight classification maps obtained using DiFuMo.
This is illustrated on \autoref{fig:decoding_coefs} for face-versus-place
decoding in the working-memory HCP study.

\section{Details on biomarker prediction}

We consider multiple datasets to account for the diversity of prediction targets
in biomarker prediction problem. We report datasets, prediction groups and
prediction targets in \autoref{tab:rest}.

\begin{table}[t!]
    \small
    \begin{center}
        \begin{tabular}{c|c|c}
            \hspace*{0.3em}\cellcolor{gray!25}\bf{Rest-fMRI} &
            \hspace*{0.2em}\cellcolor{gray!25}\bf{Prediction groups} &
            \hspace*{0.5em}\cellcolor{gray!25}\bf{Samples} \\
            \hline\\[-3.8mm]
            \multirow{2}{*}{HCP900} & High IQ vs Low IQ & \multirow{2}{*}{443
            subjects} \\
                   & \cellcolor{gray!25}213/230  & \\
            \hline\\[-3.8mm]
            \multirow{2}{*}{ABIDE}  & Autism vs control & \multirow{2}{*}{866
            subjects} \\
                   & \cellcolor{gray!25}402/464  & \\
            \hline\\[-3.8mm]
            \multirow{2}{*}{ACPI}   & Marijuana use vs control &
            \multirow{2}{*}{126 subjects} \\
                   & \cellcolor{gray!25}62/64  & \\
            \hline\\[-3.8mm]
            \multirow{2}{*}{ADNI}  & Alzheimers vs MCI & \multirow{2}{*}{136
            subjects} \\
                   & \cellcolor{gray!25}40/96  & \\
            \hline\\[-3.8mm]
            \multirow{2}{*}{ADNIDOD}  & PTSD vs control & \multirow{2}{*}{167
            subjects} \\
                     & \cellcolor{gray!25}89/78 & \\
            \hline\\[-3.8mm]
            \multirow{2}{*}{COBRE}  & Schizophrenia vs control &
            \multirow{2}{*}{142 subjects} \\
                     & \cellcolor{gray!25}65/77 & \\
            \hline\\[-3.8mm]
            \multirow{2}{*}{CamCAN}  & Age & \multirow{2}{*}{626 subjects} \\
                    & \cellcolor{gray!25}$24 - 86$ & \\
        \end{tabular}
    \end{center}
        \caption{\textbf{Resting-state fMRI datasets used in the
        pipeline described on \autoref{sec:rsfmri-pipeline} for predicting
        phenotypic labels from functional connectomes.}
        In CamCAN, age is predicted using ridge regression. The groups from other datasets are predicted using logistic regression. IQ - Fluid intelligence,
        PTSD - Post Traumatic Stress Disorder, MCI - Mild Cognitive Impairment.
        \label{tab:rest}}
\end{table}

\subsection{Input data and prediction settings}
The connectivity features built from functional atlases predict
various clinical outcomes (neuro-degenerative and neuro-psychiatric disorders,
drug abuse impact) and psychological traits.

\paragraph{Group classification}

We use the Alzheimer's
Disease Neuroimaging Initiative\footnote{\url{www.adni-info.org}}
(ADNI) and (ADNIDOD) \citep{mueller2005} to predict neuro-degenerative diseases. We discriminate between
Alzheimer's Disease (AD) from Mild Cognitive Impairment (MCI) group
on ADNI. We discriminate between post-traumatic stress disorder
(PTSD) and healthy individuals on ADNIDOD.
We use data from
the Center for Biomedical Research Excellence\footnote{\url{https://www.mrn.org/research/details/cobre}}
(COBRE \cite{calhoun2012} to predict schizophrenia diagnosis of individuals. We classify autism and healthy individuals on
Autism Brain Imaging Data Exchange database (ABIDE, \cite{dimartino2014autism},
Finally, we consider data from Addiction Connectome Preprocessed
Initiative\footnote{\url{http://fcon_1000.projects.nitrc.org/indi/ACPI/html/}}
(ACPI), where we discriminate Marijuana consumers versus control subjects.

\begin{figure}
    \centerline{%
        \includegraphics[width=1.\linewidth]{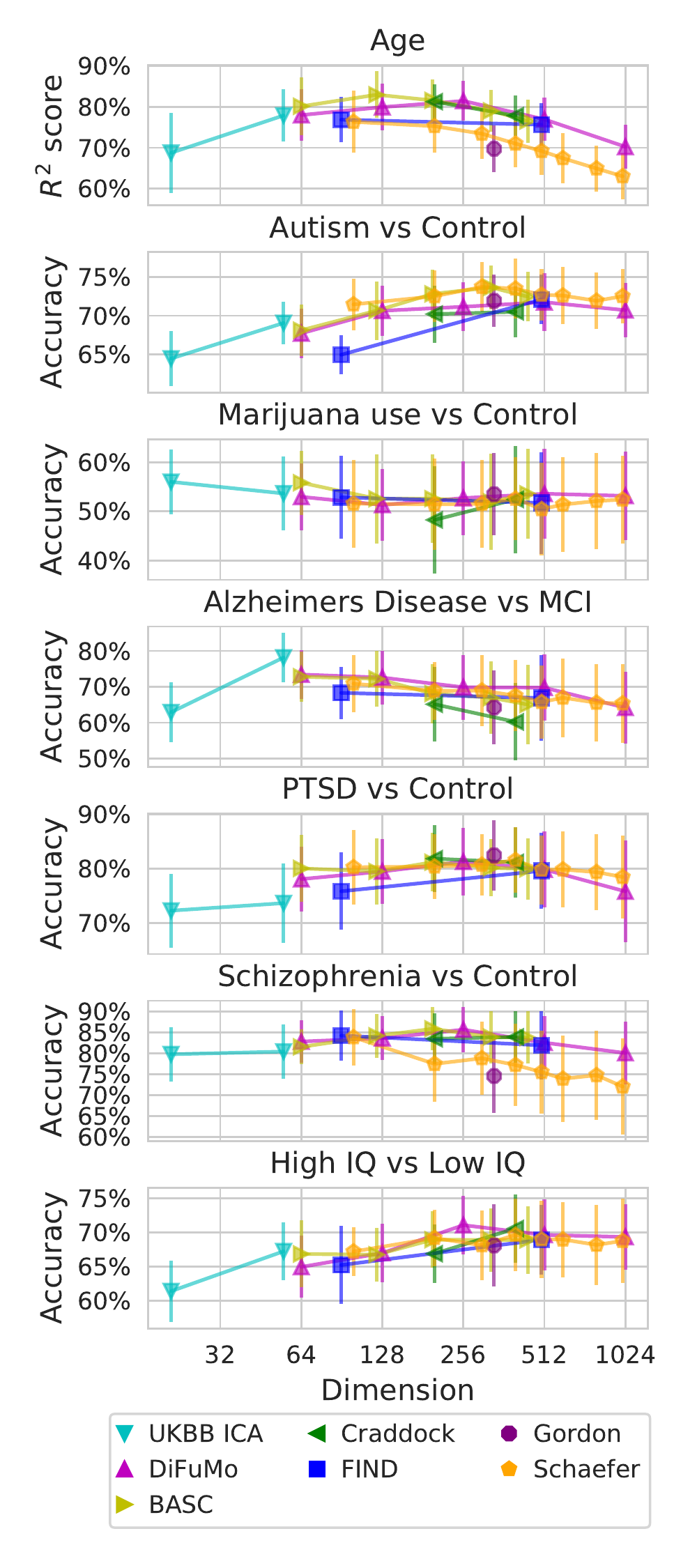}%
   \raisebox{130ex}{%
       \llap{\rlap{\textbf{\sffamily Predicting phenotypes from functional connectomes}}\hspace*{0.9\linewidth}}%
   }%
    }%
    \caption{\textbf{Connectome prediction scores for each brain atlases and target}:
        Each marker denotes the mean performance of using a certain brain atlas; error bars are the standard deviation
        of the prediction scores for this atlas. BASC and DiFuMo-based
        atlases give good prediction scores up to $k=256$ ROIs.
    \label{fig:decoding_rest_error_bars}
    }
\end{figure}

\paragraph{Psychological traits} We first stratify
individuals from HCP900 release \citep{vanessen2013hcp} into groups of high and low IQ, and perform binary classification on these. The details
about the stratification into these groups are described in \cite{dadi2019}.

\paragraph{Age regression}
We use
Cambridge Center for Ageing and Neuroscience (CamCAN) dataset
\citep{taylor_2017} to study brain ageing. This study comprises wide range of age groups spanning from
24 -- 86. We use ridge regression to predict age on this dataset.

\subsection{Data acquisition parameters and pre-processing pipelines}
The data acquisition details for ADNI, ADNIDOD, COBRE, ABIDE, ACPI and
HCP are described in \cite{dadi2019}; those for CamCAN in \cite{taylor_2017}.
We pre-process individuals from CamCAN, ADNI, ADNIDOD and
COBRE. All rs-fMRI acquistions are pre-processed with standard steps,
described in \cite{dadi2019}.
The other considered datasets provide preprocessed data.
We report the total number of subjects included in the
analysis in \autoref{tab:rest}, after excluding for severe scanning artifacts, head movements with
amplitude larger than 2~mm and individuals who have more than one
clinical diagnosis,

\paragraph{Confound removal and temporal signal pre-processing}
The strategy we use for cleaning temporal signals is the same as in
\cite{dadi2019}. We brieftly outline these steps here. We regress out 10 CompCor
\citep{behzadi2007} components on the whole brain and six motion related signals
which are provided in the ADNI, ADNIDOD, COBRE, CamCAN datasets. We
do not perform any additionnal preprocessing steps on ABIDE, ACPI and HCP. For all datasets, the signal of each region is normalized, detrended
and bandpass-filtered between 0.01 and 0.1Hz.  All these steps are done with \textit{nilearn}
\citep{abraham2014}.

\subsection{Detailed results}

\autoref{fig:relative_median_rest} summarizes the
impact of the brain atlases and ROIs in predicting diverse targets on
rs-fMRI images. \autoref{fig:decoding_rest_error_bars}
shows the absolute prediction scores for each target separately. DiFuMo-based predictions are on par with those using UKBB ICA components, \citet{craddock2012} and BASC atlases.

\section{Intra-subject encoding}
\label{sec:intra_subject_analysis}
In \autoref{sec:glm-pipeline}, we compare group-level $z$-maps computed at the
voxel-level and on reduced data using the DICE similarity coefficient. We also
performed an intra-subject, across sessions, \textit{standard} analysis. We
consider the Rapid-Serial-Visual-Presentation (RSVP) language task of Individual
Brain Charting (IBC) (see \cite{pinho2018ibc} for details on experimental
protocol and data pre-processing).

\begin{figure}[t!]
    \begin{minipage}{1.\linewidth}%
    \includegraphics[width=0.53\linewidth]{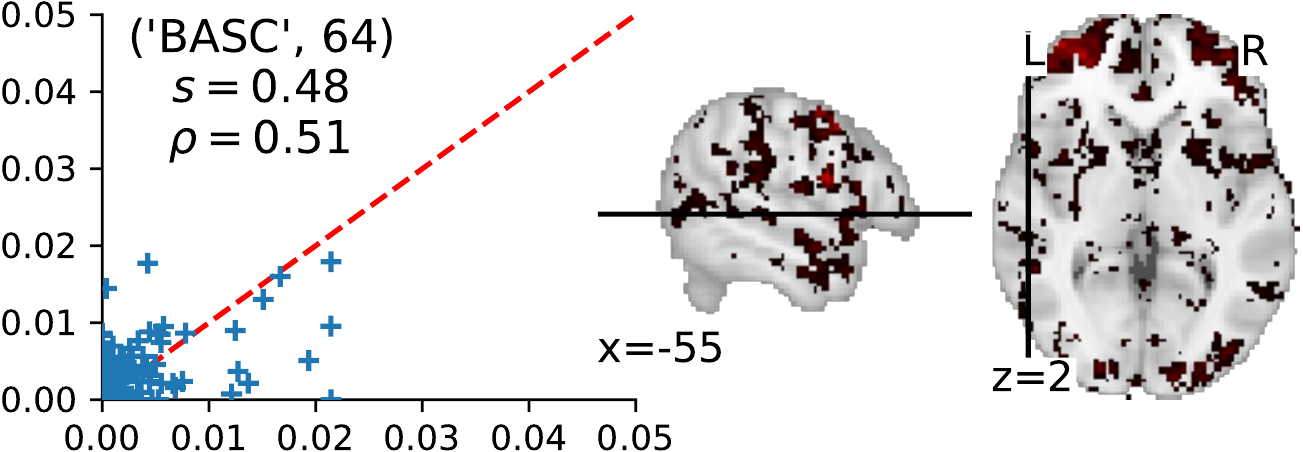}%
    \hspace*{0.001\linewidth}
    \includegraphics[width=0.53\linewidth]{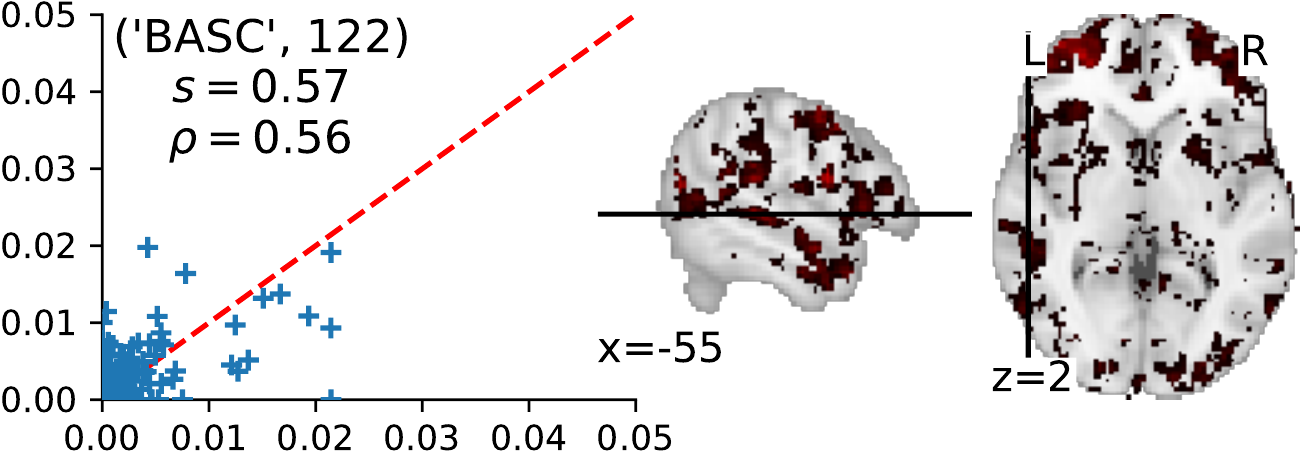}%
    \raisebox{14ex}{%
    \llap{\rlap{\textbf{\sffamily Encoding activations across single-subject sessions}}\hspace*{0.99\linewidth}}%
    }%
    \end{minipage}%
    \hfill%
    \vspace*{0.05\linewidth}
    \begin{minipage}{1.\linewidth}%
    \includegraphics[width=0.53\linewidth]{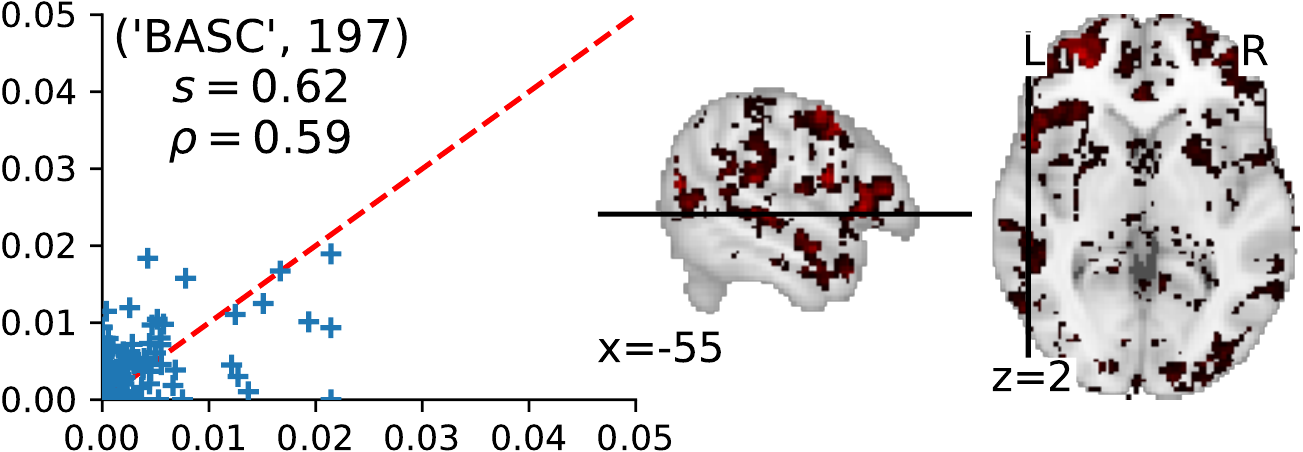}%
    \hspace*{0.001\linewidth}
    \includegraphics[width=0.53\linewidth]{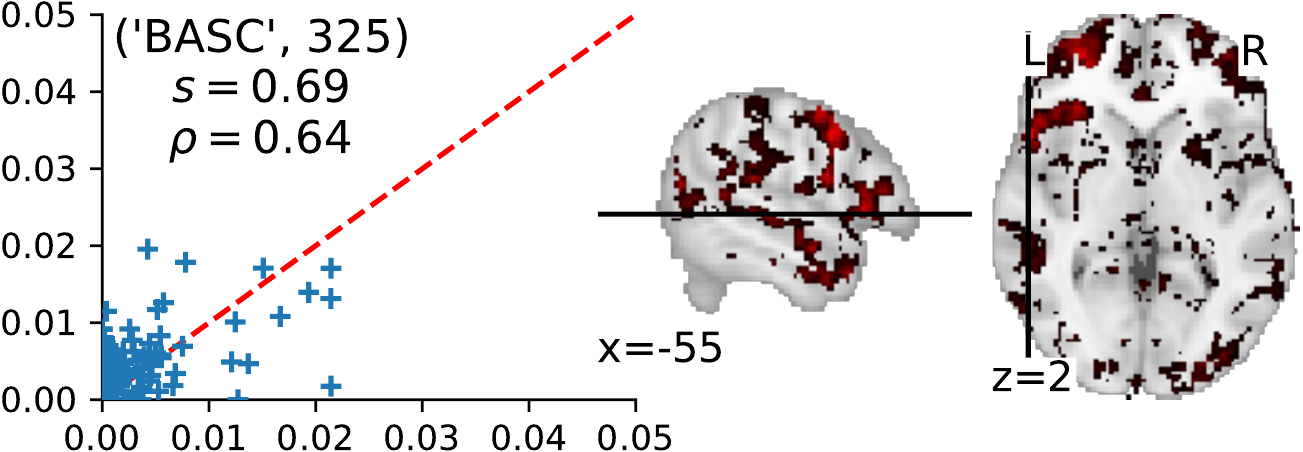}%
    \end{minipage}%
    \hfill%
    \vspace*{0.05\linewidth}
    \begin{minipage}{1.\linewidth}%
    \includegraphics[width=0.53\linewidth]{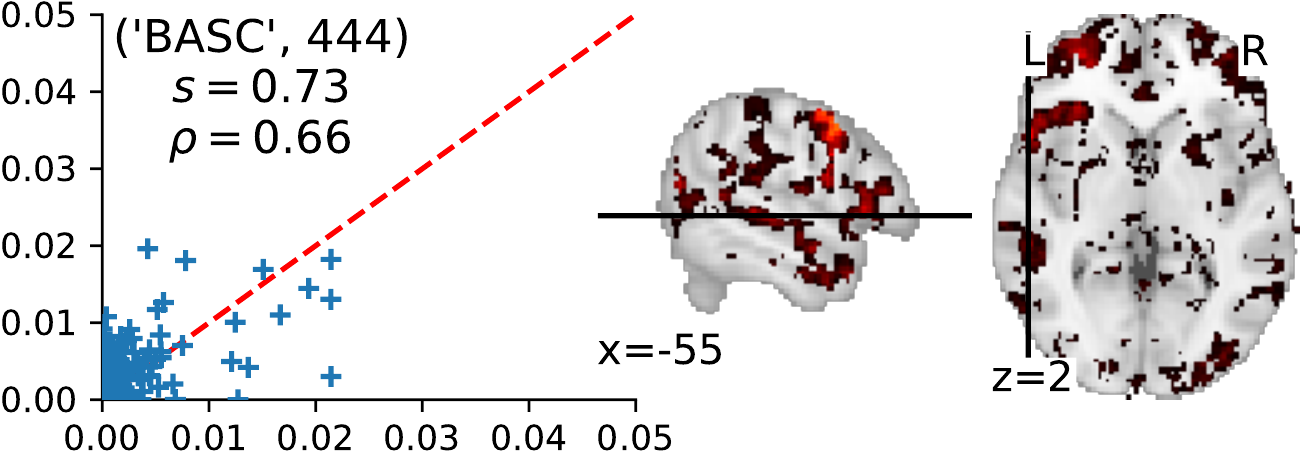}%
    \hspace*{0.001\linewidth}
    \includegraphics[width=0.53\linewidth]{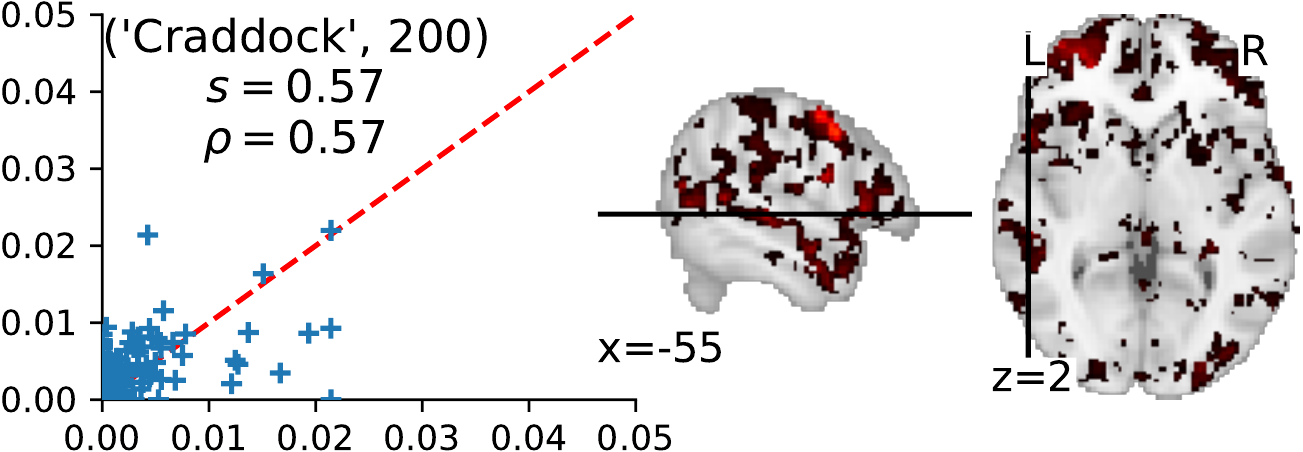}%
    \end{minipage}%
    \hfill%
    \vspace*{0.05\linewidth}
    \begin{minipage}{1.\linewidth}%
    \includegraphics[width=0.53\linewidth]{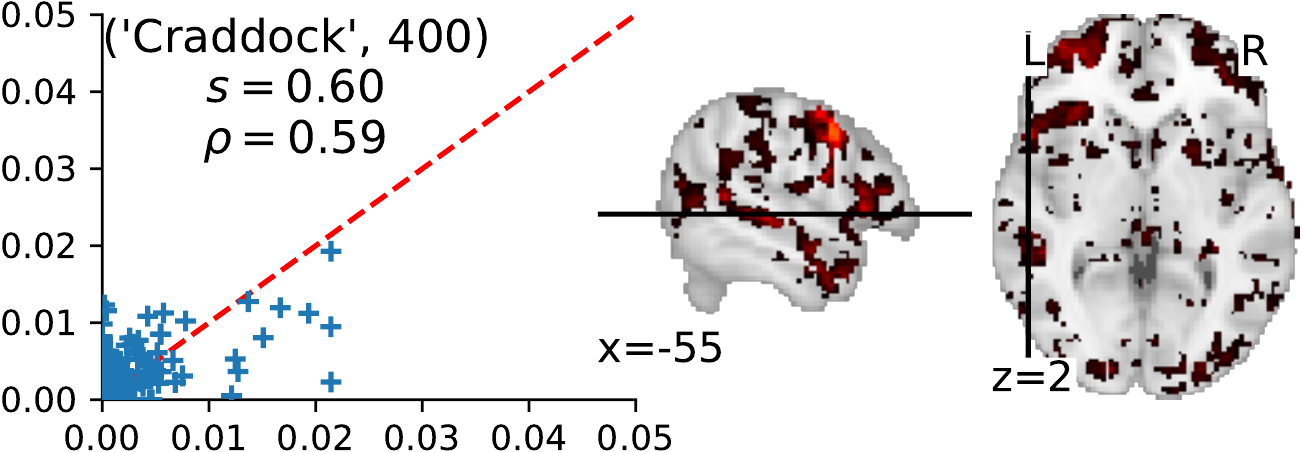}%
    \hspace*{0.001\linewidth}
    \includegraphics[width=0.53\linewidth]{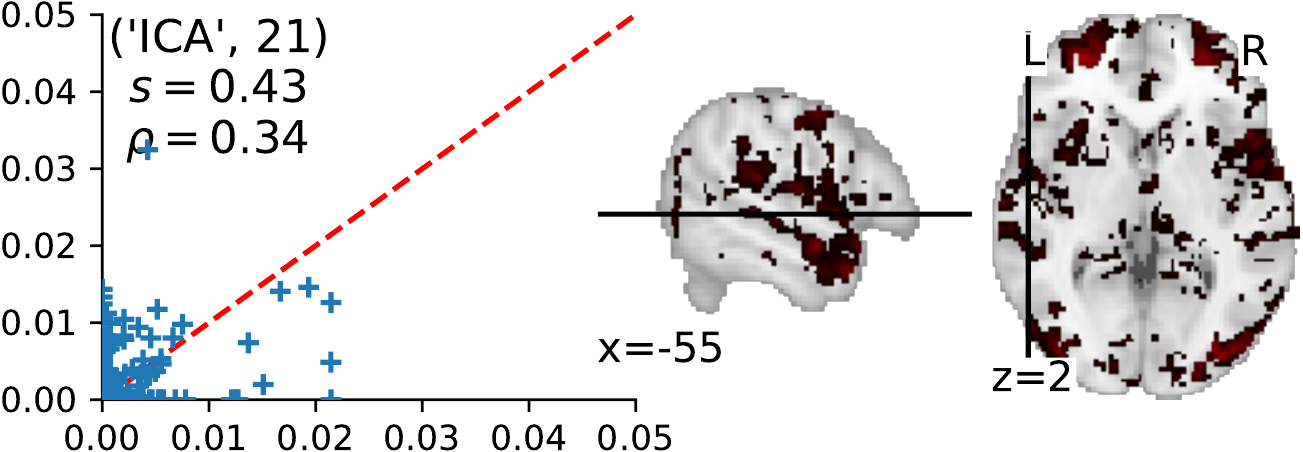}%
    \end{minipage}%
    \hfill%
    \vspace*{0.05\linewidth}
    \begin{minipage}{1.\linewidth}%
    \includegraphics[width=0.53\linewidth]{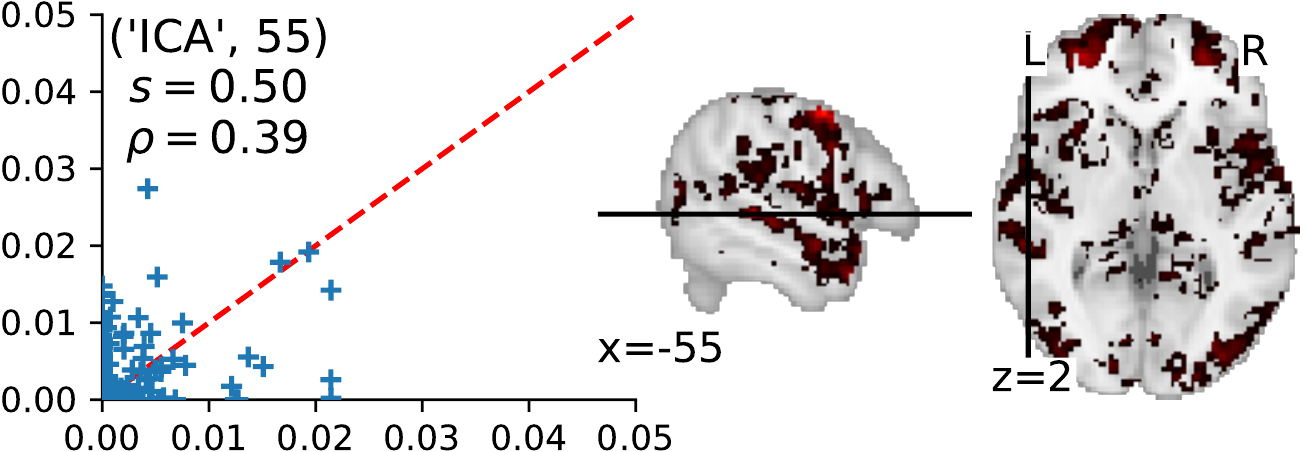}%
    \hspace*{0.001\linewidth}
    \includegraphics[width=0.53\linewidth]{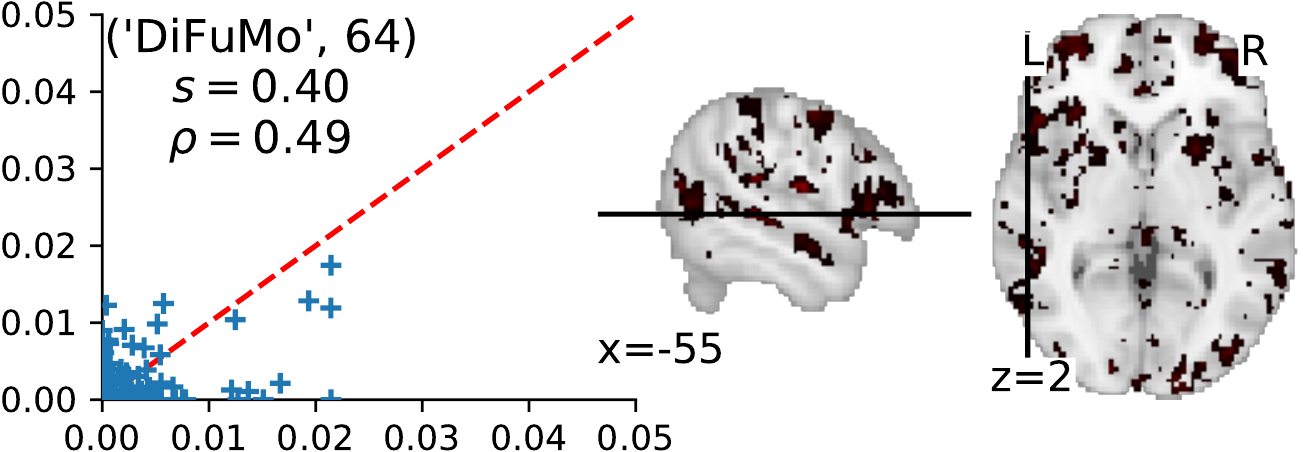}%
    \end{minipage}%
    \hfill%
    \vspace*{0.05\linewidth}
    \begin{minipage}{1.\linewidth}%
    \includegraphics[width=0.53\linewidth]{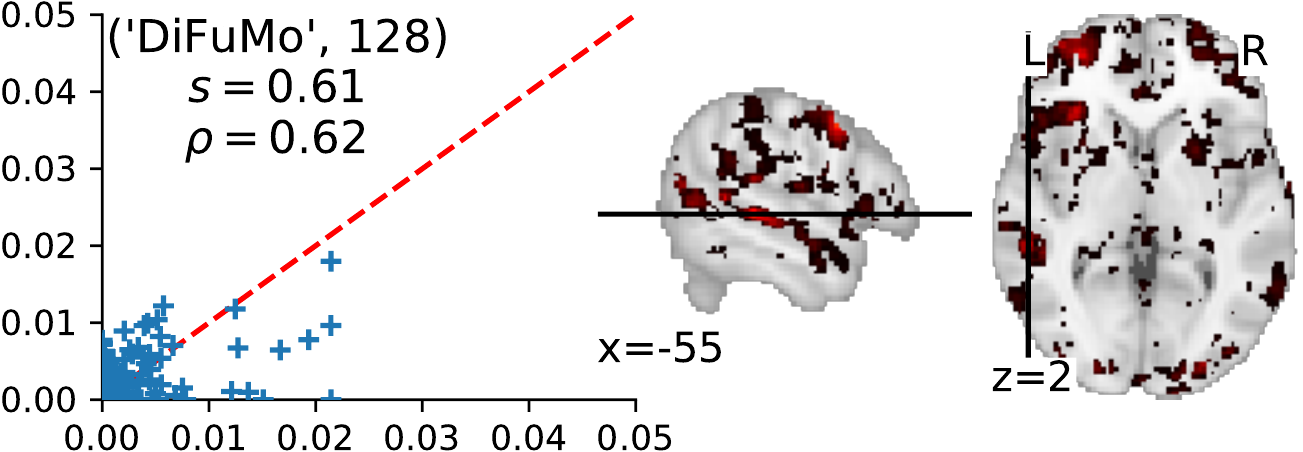}%
    \hspace*{0.001\linewidth}
    \includegraphics[width=0.53\linewidth]{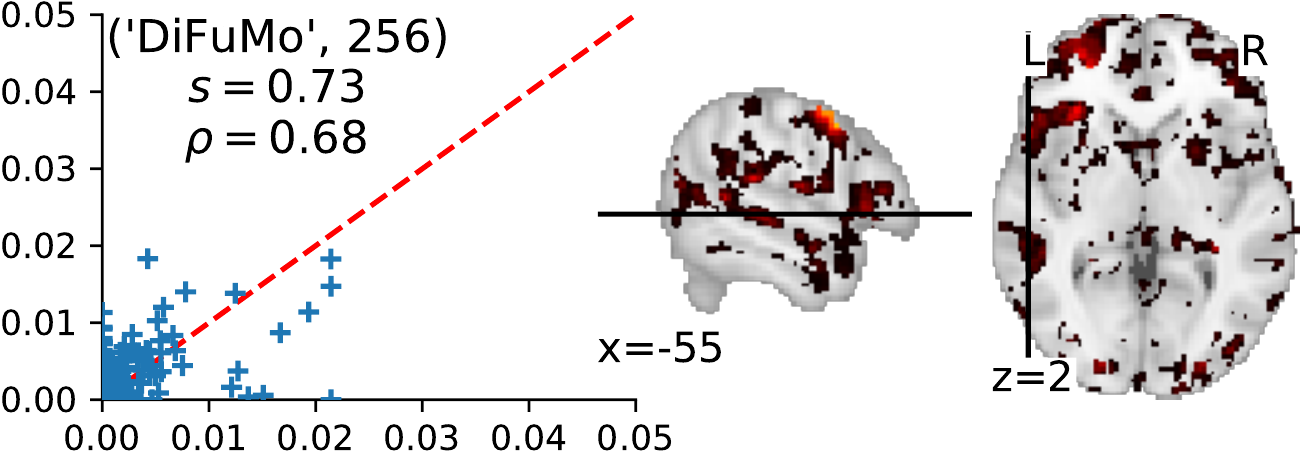}%
    \end{minipage}%
    \hfill%
    \vspace*{0.05\linewidth}
    \begin{minipage}{1.\linewidth}%
    \includegraphics[width=0.53\linewidth]{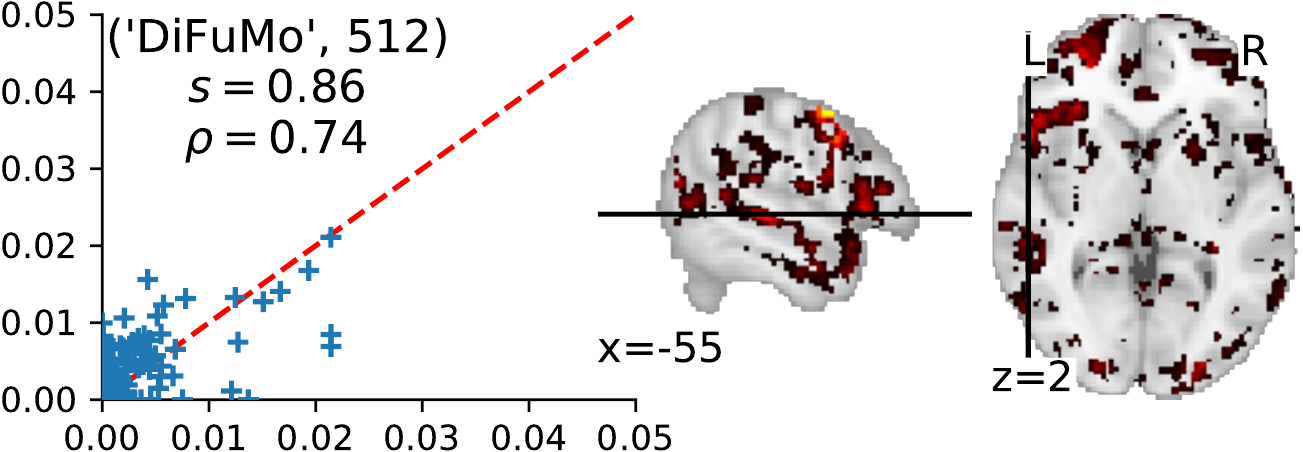}%
    \hspace*{0.001\linewidth}
    \includegraphics[width=0.53\linewidth]{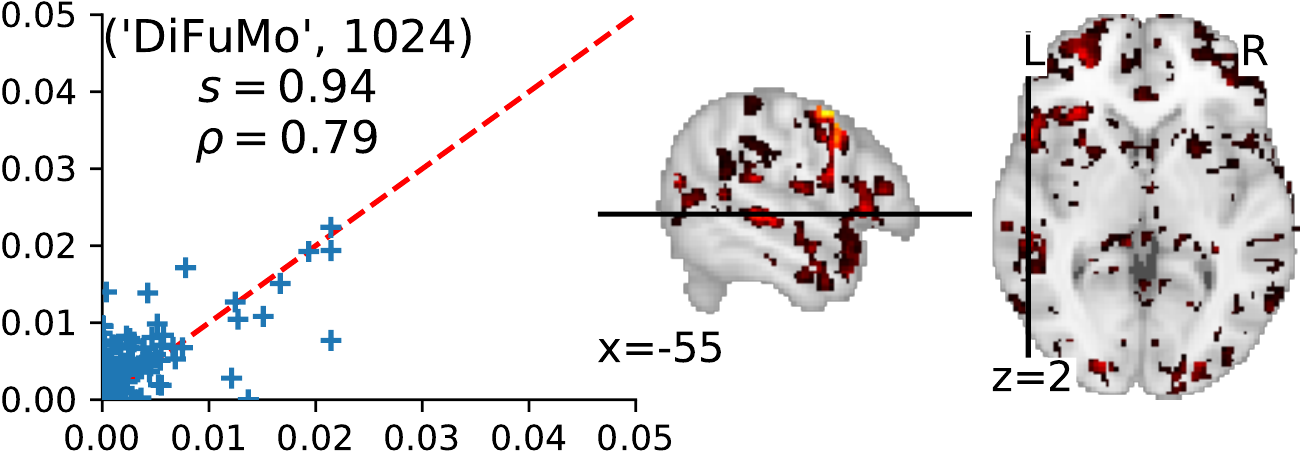}%
    \end{minipage}%
    \hfill%
    \vspace*{0.05\linewidth}
    \begin{minipage}{1.\linewidth}%
    \includegraphics[width=0.58\linewidth]{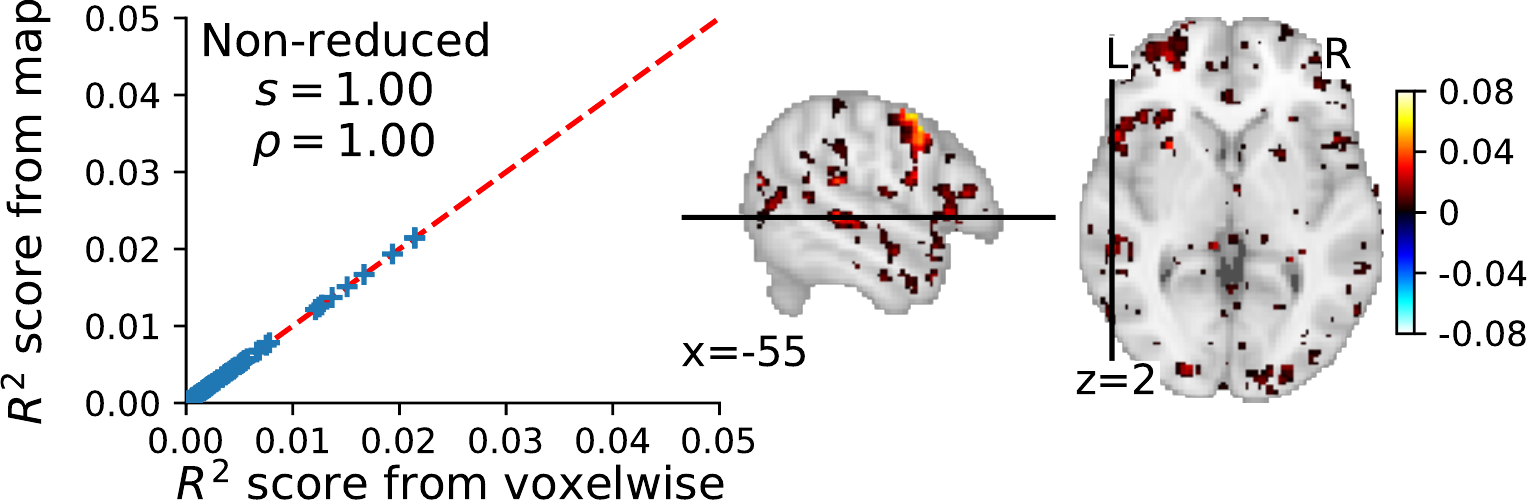}%
    \hspace*{0.01\linewidth}
    \end{minipage}%
    \caption{\textbf{Intra-subject univariate prediction of brain response
    in the language task protocol of the IBC dataset.}
    We compare $R^2$-maps obtained using voxel based and functional-atlas based encoding models. Encoding models based on high-order atlases better explain the variance of an unseen session.
     The comparison is made for a single subject; results are similar across subjets.}
    \label{fig:encoding_r2}
    \end{figure}

\paragraph{Encoding model} In this setting, we fit a GLM on the several acquisition sessions of each
subject considered separately. That is, we compute a single $\beta$-map per
session and condition, forming a set of maps $\bbe \in \RR^{q \times p}$. $\bbe$
is either computed directly at the voxel-level or using functional atlases, in
which case we set $\bbe = \bbe_{red} \D^\top$, with $\bbe \in \RR^{q \times k}$.

We then use a leave-one-session-out cross-validation scheme to compare the observed,
single-session, time series $\Y \in \RR^{n \times p}$ to the reconstructed
time-series $\hat \Y = \X \bar \bbe$, where $\bar \bbe$ are the average
$\beta$-maps across the $5$ training sessions. We obtain $R^2$-maps, where each voxel holds the proportion of variance explained by the model
\begin{equation}
    r_{i} = 1 - \frac{\Vert \y_i - \hat \y_i \Vert_2^2}{\Vert \y_i - \bar y_i \Vert_2^2},
\end{equation}
where $\y_i$ is the univariate time-series in $\RR^n$ associated to voxel $i$
and $\bar y_i$ is its temporal mean. We finally average $R^2$ scores across
leave-one-session-out folds, and threshold non-positive values. The resulting
$R^2$-maps provides information on how much encoded $\beta$-maps are able to
predict univariate voxel activation on new sessions. A value close to $1$ means
that the voxel activation is well predicted by the encoding model, while a $0$
value means that the voxel activation cannot be predicted. We compare the
$R^2$-maps across the various data-reduction methods for estimating~$\bbe$. 

\paragraph{Validation} To measure the difference between $R^2$ maps $\R$
computed from voxels and $R^2$ maps $\tilde \R$ computed from DiFuMos, we report
correlation coefficients $\rho$ between $\R$ and $\tilde \R$, and the slope $s$
predicting the activations $\tilde \R$ from the activations $\R$. This slope
indicates a form of signal loss due to using functional atlases. We
expect it to be smaller than $1$, in part because projection on functional atlases have a
noise reduction effect.

\paragraph{Results} \autoref{fig:encoding_r2}, using higher order DiFumo
atlases leads to a loss of
explained variance $R^2$ of only $6 \%$ compared to working directly with
voxels, which may imputed to a denoising effect. Qualitatively, the $R^2$ maps
are much comparable. DiFuMo ($k = 1024$) is therefore suitable for intra-subject
encoding tasks; they make these much less costly. Using lower-order atlases
yield stronger signal loss.

\section{Extra meta-analysis maps}
\label{app:meta-analysis_more_topics}
\begin{figure*}[t!]
    \hspace*{-.05\linewidth}%
    \includegraphics[width=1.05\linewidth]{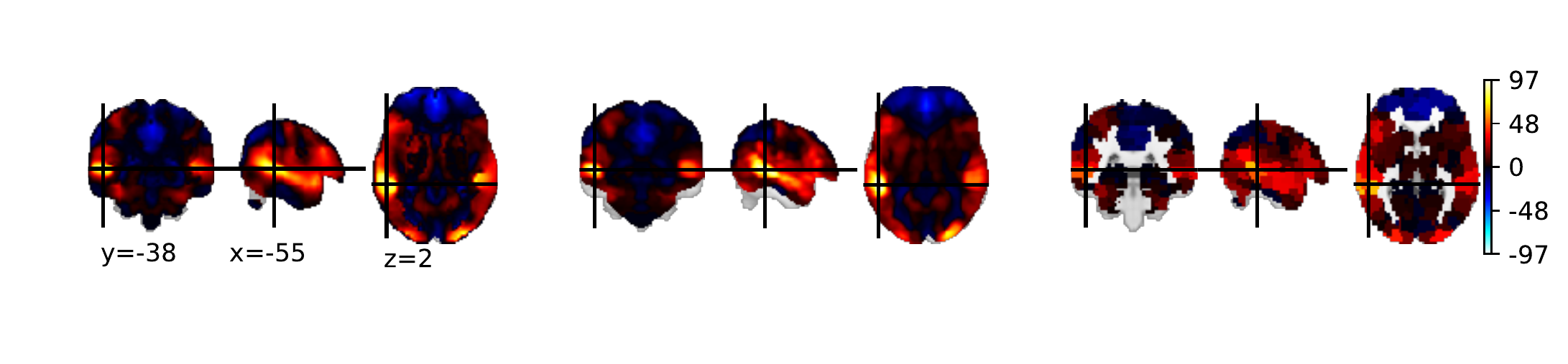}%
    \raisebox{23.5ex}{%
        \llap{\rlap{\textbf{\sffamily a. Meta-analysis of "language"}}
    \hspace*{0.96\linewidth}}%
    }%
    \vspace*{-7.7ex}

    \hspace*{-.05\linewidth}%
    \includegraphics[trim={0 0 0 1cm}, clip, width=1.05\linewidth]{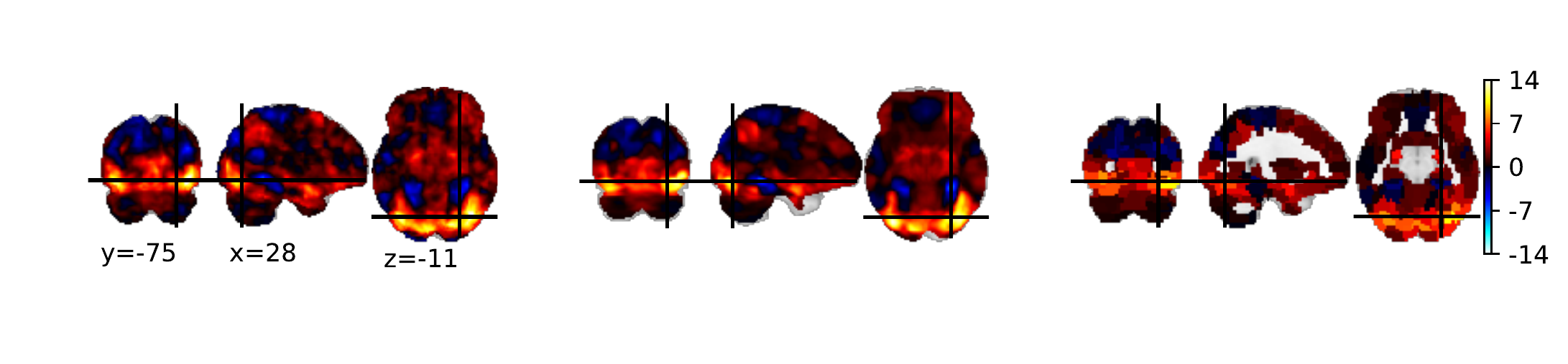}%
    \raisebox{5ex}{%
        \llap{\rlap{\textbf{\sffamily b. Meta-analysis of "face"}}
    \hspace*{0.96\linewidth}}%
    }%
    \llap{\raisebox{23ex}{\parbox{\linewidth}{%
    \sffamily \quad\qquad Non-reduced images\quad\hfill Image reduced with DiFuMo
    \hfill Image reduced with BASC \qquad\quad\vbox{}%
    }}}\vspace*{1ex}
    \caption{\textbf{Meta-analysis on cognitive topics --language (a.) and
            face (b.) -- from
            statistical images}: We
            compare images reconstructed with DiFuMo ($k=1024$) and
            BASC ($k=444$) with voxel-level averages (right). The topic-related activations
            are better visualized using DiFuMo (middle) than using BASC (left). DiFuMo results are closer to voxel-level averages, as the signal loss is minimal when projecting on this atlas.}
            \label{fig:meta-analysis_more_cognitive_topics}
\end{figure*}

\autoref{fig:meta-analysis_more_cognitive_topics} shows the meta-analysis
summary images for two additional cognitive topics: \textit{language} and
\textit{face}. We compare non-reduced images with reduced images using DiFuMo
($k=1024$) and BASC ($k=444$). The images reduced with DiFuMos are easier to
interpret than the ones reduced with BASC for both topics.
Quantitatively, we recall that \autoref{fig:compression} shows the better
performance of DiFuMos for image compression.

\section{DiFuMos naming details}
\label{app:difumo_labeling}
A measure of overlap with a reference anatomical atlas allows to match each
DiFuMo component with a specific anatomical region, e.g. ``postcentral gyrus''.
Where there are more than one component for each anatomical region, the
functional atlas region are further characterized by an anatomical spatial
descriptions, e.g. ``postcentral gyrus inferior''. Finally, we append the
localisation of the region in the left or right hemisphere, e.g.
``postcentral gyrus inferior RH''.
Some of the nodes from DiFuMo atlases overlaps a fraction of several regions
in the anatomical atlas---those are named by a trained neuroanatomist.

\begin{figure*}[t]
    \hspace*{.01\linewidth}%
    \includegraphics[width=1.\linewidth]{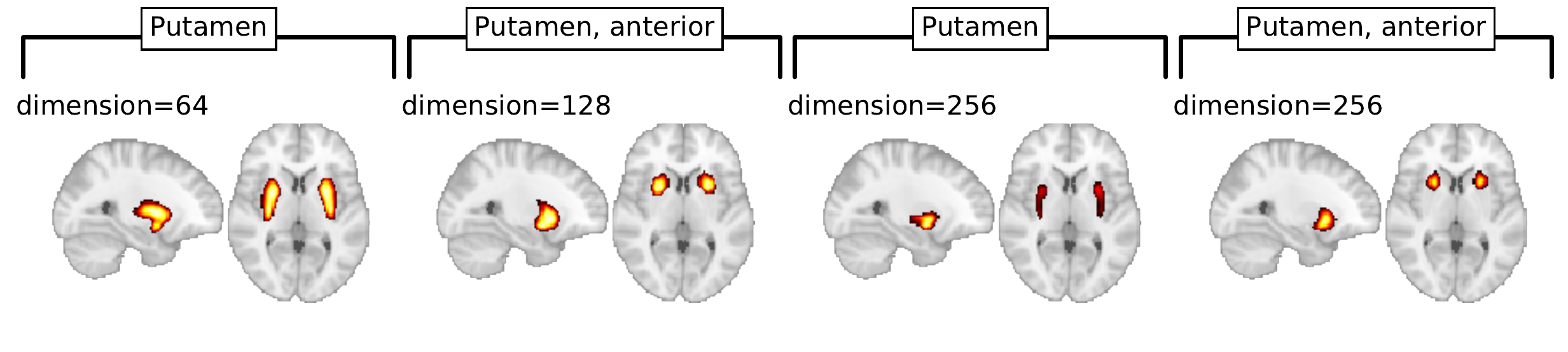}%
    \raisebox{27.5ex}{%
        \llap{\rlap{\textbf{\sffamily Modes around putamen: a. Smaller atlases contains bilateral networks}}
    \hspace*{0.96\linewidth}}%
    }%
    \vspace*{-2.6ex}

    \hspace*{.01\linewidth}%
    \includegraphics[width=1.\linewidth]{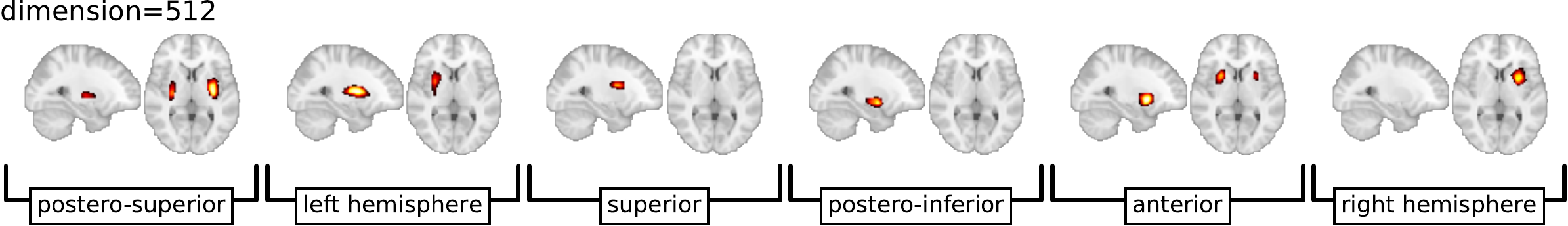}%
    \raisebox{19ex}{%
        \llap{\rlap{\textbf{\sffamily b. Increasing the atlas dimension splits those networks into different components}}
    \hspace*{0.96\linewidth}}%
    }%
    \vspace*{-1.1ex}
    \caption{\textbf{Interpretation of higher-dimensional modes of
     DiFuMo}: The putamen segmentation is refined as dimension of DiFuMos
     increases. A single mode contain the left and right putamen in lower
     dimension \textbf{(a)}, when higher order atlases holds separate components
     for them. Larger atlases model the detailed
     organization within the sub-structures, which may be crucial in discriminative
     tasks.}
    \label{fig:putamen_supp}
\end{figure*}


\begin{table*}[b]
    \centering%
\footnotesize%
\begin{tabular}{ccccccl}
    \hline
    {fMRI study}   & {Version}  & {Cognitive task}  &  \rotatebox{75}{\#Subjects} &
    \rotatebox{75}{\#Sessions} &  \rotatebox{75}{\#Runs} & {Conditions} \\
   \hline
   \citep{schonberg2012} & ds000001\_R2.0.4 & Balloon Analog   & $16$      & \_         & $3$    &  balloon analog risk    \\
            &                  &  Risk-taking     &           &            &        &                         \\
   \rowcolor{gray!13}
   \citep{aron2006} & ds000002\_R2.0.5 & Classification learning & $17$ & \_ & $2$ & deterministic classification\\
   \rowcolor{gray!13}
            &                  &                         &     &    &   & mixed event related probe\\
   \rowcolor{gray!13}
            &                  &                         &   &  &  &  probabilistic classification\\
   \citep{xue2007} & ds000003\_R2.0.2 & Rhyme judgment          & $13$  &  \_&  &  rhyme judgment\\
   \rowcolor{gray!13}
   \citep{jimura2014} & ds000006\_R2.0.1 & ds000006                & $14$ &  $2$ &  $6$  & living nonliving decision-   \\
   \rowcolor{gray!13}
            &                  &                         &      &      &       & with plain or mirror reversed text  \\
   \citep{xue2008} & ds000007\_R2.0.1 & Stop-signal task with   & $20$ &  \_    & $2$  & stop manual\\
            &                  & spoken \& manual responses  &      &      &   & stop vocal\\
            &                  &                             &      &      &  & stop word\\
   \rowcolor{gray!13}
   \citep{aron2007} & ds000008\_R2.0.0 & Stop-signal task with & $14$ & \_ & $3$ &   conditional stop signal\\
   \rowcolor{gray!13}
            &                  & unconditional and conditional & & & & stop signal\\
   \rowcolor{gray!13}
            &                  & stopping & & & &\\
   \citep{foerde2006} & ds000011\_R2.0.1 & Classification learning & $14$ & \_ & $2$ &  Classification probe without  \\
            &                  & and tone counting       &      &     &     &  feedback  \\
            &                  &                         & & & &  Dual task weather prediction \\
            &                  &                   & & & &  Single task weather prediction \\
            &                  &                   & & & &  Tone counting \\
   \rowcolor{gray!13}
   \citep{rizk2011} & ds000017\_R2.0.1 & Classification learning & $8$ & 2 & 3  & probabilistic classification\\
   \rowcolor{gray!13}
            &                  & and stop-signal (1 year test-retest) & & &  &  selective stop signal task\\
   \citep{alvarez2011} & ds000051\_R2.0.2 & Cross-language & $13$ & \_  & $8$  &  abstract concrete judgment\\
            &                 & repetition priming & & & &\\
   \rowcolor{gray!13}
   \citep{poldrack2001} & ds000052\_R2.0.0 & Classification learning & $14$ & \_ & $2$ &  weather prediction\\
   \rowcolor{gray!13}
            &                 & and reversal             &      &    &     &  reversal weather prediction\\
   \citep{kelly2016} & ds000101\_R2.0.0 & Simon task              & $21$ &  \_ & $2$ &  simon\\
   \rowcolor{gray!13}
   \citep{kelly2008} & ds000102\_R2.0.0 & Flanker task & $26$ & \_ & $2$ & flanker\\
   \rowcolor{gray!13}
            &                  &  (event-related) & & & &\\
   \citep{haxby2001} &ds000105\_R2.0.2  & Visual object recognition & $6$ & \_ & $12$ &  object viewing\\
   \citep{otoole2005} &                 &                           &     &  &  &  \\
   \citep{hanson2004} &                 &                           &     &  &  &  \\
   \rowcolor{gray!13}
   \citep{duncan2009consistency} & ds000107\_R2.0.2 & Word and  & $49$ &\_ & $2$ & 1-back task\\
   \rowcolor{gray!13}
            &                  & object processing & & & &\\
   \citep{moran2012} & ds000109\_R2.0.2 & False belief task & $36$ & \_ & $2$ & theory of mind\\
   \rowcolor{gray!13}
   \citep{uncapher2011} & ds000110\_R2.0.1 & Incidental encoding task & $18$ & \_ & $10$ & Incidental encoding task\\
   \rowcolor{gray!13}
            &                  & (Posner Cueing Paradigm) &  &  &   & \\
   \citep{gorgolewski2013} & ds000114\_R2.0.1 & A test-retest fMRI dataset  & $10$ & $2$ & \_  & covert verb generation\\
            &                  & for motor, language and &      &     &     & finger footlips\\
            &                  & spatial attention functions &      &     &     & line bisection\\
            &                  &                             &      &     &     & overt verb generation\\
            &                  &                             &      &     &     & overt word generation\\
   \rowcolor{gray!13}
   \citep{repovs2012} & ds000115\_R2.0.0 & Working memory in healthy & $1$ & \_ & \_ & letter 0-back task\\
   \rowcolor{gray!13}
            &                  & and schizophrenic individuals & & & &  letter 1-back task\\
   \rowcolor{gray!13}
            &                  &                               & & & &  letter 2-back task\\
   \citep{nicoletta2014} & ds000133\_R1.0.0 & Modafinil alters intrinsic & $26$ & $2$ & $3$  & rest \\
            &                  & functional connectivity of the &  &   &   &  \\
            &                  & right posterior insula: a &  &  &  &  \\
            &                  & pharmacological &  &  &  &  \\
            &                  & resting state fMRI study&  &  &  &  \\
   \rowcolor{gray!13}
   \citep{timothy2014} & ds000164\_R1.0.1 & Stroop task & $28$ & \_ & \_ & stroop \\
   \citep{gabitov2015} & ds000170\_R1.0.1 & Learning and memory: motor & $15$ & \_ & $3$ & Trained Hand Trained Sequence\\
            &                  & skill consolidation and & & & & Trained Hand Untrained Sequence\\
            &                  & intermanual transfer & & & & Untrained Hand Trained Sequence\\
   \rowcolor{gray!13}
   \citep{lepping2016a} & ds000171\_R1.0.0 & Neural Processing of Emotional & $39$ & \_ &  $5$ & music\\
   \rowcolor{gray!13}
   \citep{lepping2016b}        &                  & Musical and Nonmusical & & & & non music\\
   \rowcolor{gray!13}
            &                  & Stimuli in Depression & & & &\\
   \citep{iannilli2016} & ds000200\_R1.0.0  & Pre-adolescents Exposure & $1$ & \_ & \_  &  olfactory\\
            &                   & to Manganese & & & &\\
   \rowcolor{gray!13}
   \citep{christian2017} & ds000203\_R1.0.2 & Visual imagery and & $26$ & \_ & $2$ & visual imagery-\\
   \rowcolor{gray!13}
            &                  & false memory for pictures & & & & false memory\\
   \citep{kim2016} & ds000205\_R1.0.0 & Affective Videos & $11$ & \_ & $2$ & functional localizer\\
            &                  &                  &   &  &   & view \\
   \rowcolor{gray!13}
   \citep{romaniuk2016} & ds000214\_R1.0.0 & EUPD Cyberball & $40$  & \_ & \_ & Cyberball\\
   \citep{arnab2017} & ds000220\_R1.0.0 & Cost Analysis TBI & $26$ & 2 & \_ & rest\\
   \hline
\end{tabular}

    \caption{Large-scale fMRI datasets downloaded from OpenNeuro to build
    our multi-scale functional atlases. Data are pre-processed using
    \textit{fMRIprep}. The data acquisition parameters of each study are listed
    on \autoref{tab:data_acquisition_somf_learner}. The corpus is 2.4TB in total.}
    \label{tab:task_openneuro}
\end{table*}

\begin{table*}[b]
    \centering%
\footnotesize%
\begin{tabular}{p{27ex}p{21.5ex}ccccccccc}
    \hline
\hspace*{-1.7ex}fMRI study   & MR scanner & \hspace*{-2ex}Slice     & \hspace*{-2ex}FoV    & \hspace*{-2ex}Voxel size  & \hspace*{-2ex}Matrix & \hspace*{-2ex}TR      &   \hspace*{-2ex}TE     & \hspace*{-2ex}Flip angle & \hspace*{-2ex}Number of\\
                      &                          & \hspace*{-2ex}orientation & \hspace*{-2ex}$(mm)$ &  \hspace*{-2ex}$(mm)$     & \hspace*{-2ex}size   &\hspace*{-2ex}$(msec)$ & \hspace*{-2ex}$(msec)$ & \hspace*{-2ex}$(^\circ)$ & \hspace*{-2ex}volumes \\
\hline\\[-2mm]
\hspace*{-1.7ex}\citep{schonberg2012} & 3T Siemens AG            & \hspace*{-2ex}axial   &  \hspace*{-2ex}-  & \hspace*{-2ex}$4\times4\times4$ & \hspace*{-2ex}$64\times64$& \hspace*{-2ex}$2000$  &  \hspace*{-2ex}$30$    & \hspace*{-2ex}$90$  & \hspace*{-2ex}300\\
	                  & Allegra (Erlangen, Germany) &           &                 &     &             &          &         &               &\\[1.5mm]
\rowcolor{gray!13}
\hspace*{-1.7ex}\citep{aron2006}  & 3T Siemens   & \hspace*{-2ex}-    &  \hspace*{-2ex}-  &   \hspace*{-2ex}$4\times4\times4$  &\hspace*{-2ex}$64\times64$ & \hspace*{-2ex}$2000$  & \hspace*{-2ex}$30$    &  \hspace*{-2ex}$90$    & \hspace*{-2ex}180\\
\rowcolor{gray!13}
            & Allegra &           &                 &     &             &          &         &               &\\[1.5mm]
\hspace*{-1.7ex}\citep{xue2007} & 3T Siemens & \hspace*{-2ex}-  & \hspace*{-2ex}$200$ & \hspace*{-2ex}$4\times4\times4$ & \hspace*{-2ex}$64\times64$ & \hspace*{-2ex}$2000$ & \hspace*{-2ex}$30$ &  \hspace*{-2ex}$90$    & \hspace*{-2ex}$160$\\
                                & Allegra (Iselin, NJ) &           &                 &     &             &          &         &               &\\[1.5mm]
\rowcolor{gray!13}
\hspace*{-1.7ex}\citep{jimura2014}  & 3T Siemens & \hspace*{-2ex}-    &   \hspace*{-2ex}$200$        &     \hspace*{-2ex}$4\times4\times4$ & \hspace*{-2ex}$64\times64$& \hspace*{-2ex}$2000$   & \hspace*{-2ex}$30$     & \hspace*{-2ex}$90$    &  \hspace*{-2ex}205 \\
\rowcolor{gray!13}
	                  & Allegra (Erlangen, Germany) &           &                 &     &             &          &         &               &\\[1.5mm]
\hspace*{-1.7ex}\citep{xue2008}   & 3T Siemens   &  \hspace*{-2ex}-    &    \hspace*{-2ex}200        & \hspace*{-2ex}$4\times4\times4$  &  \hspace*{-2ex}$64\times64$ & \hspace*{-2ex}2000  &   \hspace*{-2ex}30   &        \hspace*{-2ex}$90$   & \hspace*{-2ex}182 \\
                                  & Allegra      &           &                 &     &             &          &         &               &\\[1.5mm]
\rowcolor{gray!13}
\hspace*{-1.7ex}\citep{aron2007}   & 3T Siemens  &  \hspace*{-2ex}-   &  \hspace*{-2ex}$200$     &   \hspace*{-2ex}$4\times4\times4$  & \hspace*{-2ex}$64\times64$    & \hspace*{-2ex}$2000$   &  \hspace*{-2ex}$30$    & \hspace*{-2ex}$90$   & \hspace*{-2ex}176\\
\rowcolor{gray!13}
                                   & Allegra     &      &            &  & &    &      &   & \\
\hspace*{-1.7ex}\citep{foerde2006}   & 3T Siemens  &  \hspace*{-2ex}-   &  \hspace*{-2ex}$200$     &  \hspace*{-2ex}$4\times4\times4$  & \hspace*{-2ex}$64\times64$    & \hspace*{-2ex}$2000$   &  \hspace*{-2ex}$30$    & \hspace*{-2ex}-   & \hspace*{-2ex}208\\
                                   & Allegra     &      &            &  & &    &      &   & \\
\rowcolor{gray!13}
\hspace*{-1.7ex}\citep{poldrack2001}   & 3T Siemens & \hspace*{-2ex}axial  &  \hspace*{-2ex}$200$     &  \hspace*{-2ex}$5\times5\times5$  & \hspace*{-2ex}$64\times64$    & \hspace*{-2ex}$3000$   &  \hspace*{-2ex}$30$    & \hspace*{-2ex}- & \hspace*{-2ex}225 \\
\rowcolor{gray!13}
                                      & Allegra     &           &                 &     &             &          &         &               &\\
\hspace*{-1.7ex}\citep{kelly2016} & 3T Siemens &  \hspace*{-2ex}- &  \hspace*{-2ex}$192$     &  \hspace*{-2ex}$3\times3\times4$  & \hspace*{-2ex}$64\times64$    & \hspace*{-2ex}$2000$   &  \hspace*{-2ex}$30$    & \hspace*{-2ex}$80$   & \hspace*{-2ex}101 \\
                                  & Allegra     &           &                 &     &             &          &         &               &\\
\rowcolor{gray!13}
\hspace*{-1.7ex}\citep{kelly2008}   & 3T Siemens & \hspace*{-2ex}-    &  \hspace*{-2ex}$192$     &   \hspace*{-2ex}$3\times3\times4$  & \hspace*{-2ex}$64\times64$    & \hspace*{-2ex}$2000$   &  \hspace*{-2ex}$30$    & \hspace*{-2ex}$80$   & \hspace*{-2ex}146\\
\rowcolor{gray!13}
                                  & Allegra     &           &                 &     &             &          &         &               &\\
\hspace*{-1.7ex}\citep{haxby2001}  & 3T GE   &  \hspace*{-2ex}sagittal  &  \hspace*{-2ex}$240$     &    \hspace*{-2ex}$3.5\times3.5\times3.5$  & \hspace*{-2ex}-   & \hspace*{-2ex}$2500$   &  \hspace*{-2ex}$30$    & \hspace*{-2ex}$90$   & \hspace*{-2ex}121\\
\rowcolor{gray!13}
\hspace*{-1.7ex}\citep{duncan2009consistency} & 1.5T Siemens & \hspace*{-2ex}-    &  \hspace*{-2ex}$192$     &  \hspace*{-2ex}$3\times3\times3.$  & \hspace*{-2ex}$64\times64$    & \hspace*{-2ex}$3000$   &  \hspace*{-2ex}$50$    & \hspace*{-2ex}- & \hspace*{-2ex}165\\
\hspace*{-1.7ex}\citep{moran2012} & 3T Siemens   &  \hspace*{-2ex}axial  & \hspace*{-2ex}-    &   \hspace*{-2ex}$3\times3\times3$  & \hspace*{-2ex}-   & \hspace*{-2ex}$2000$   &  \hspace*{-2ex}$35$    & \hspace*{-2ex}-  & \hspace*{-2ex}179\\
                                  & Tim Trio    &           &                 &     &             &          &         &               &\\
\rowcolor{gray!13}
\hspace*{-1.7ex}\citep{uncapher2011} & 3T GE Signa &  \hspace*{-2ex}axial    &  \hspace*{-2ex}-     &   \hspace*{-2ex}$3.44\times3.44\times3.44$  & \hspace*{-2ex}$64\times64$    & \hspace*{-2ex}$2000$   &  \hspace*{-2ex}$30$    & \hspace*{-2ex}$75$   & \hspace*{-2ex}186\\
\hspace*{-1.7ex}\citep{gorgolewski2013}  & 1.5T GE Signa &  \hspace*{-2ex}-   &  \hspace*{-2ex}$256$     &   \hspace*{-2ex}$4\times4\times4$  & \hspace*{-2ex}$64\times64$    & \hspace*{-2ex}$2500$   &  \hspace*{-2ex}$50$    & \hspace*{-2ex}$90$   & \hspace*{-2ex}varied \\
\rowcolor{gray!13}
\hspace*{-1.7ex}\citep{repovs2012} & 3T Tim Trio &  \hspace*{-2ex}-   &  \hspace*{-2ex}$256$     &   \hspace*{-2ex}$4\times4\times4$  & \hspace*{-2ex}$64\times64$    & \hspace*{-2ex}$2500$   &  \hspace*{-2ex}$27$    & \hspace*{-2ex}$90$   & \hspace*{-2ex}137 \\
\hspace*{-1.7ex}\citep{nicoletta2014} & 3T Philips &  \hspace*{-2ex}transaxial &  \hspace*{-2ex}$256$     &  \hspace*{-2ex}$4\times4\times4$  & \hspace*{-2ex}$64\times64$    & \hspace*{-2ex}$1671$   &  \hspace*{-2ex}$35$    & \hspace*{-2ex}$75$   & \hspace*{-2ex}145 \\
\rowcolor{gray!13}
\hspace*{-1.7ex}\citep{timothy2014} & 3T Siemens &  \hspace*{-2ex}- &  \hspace*{-2ex}-  &   \hspace*{-2ex}$3.2\times3.2\times4$  & \hspace*{-2ex}- & \hspace*{-2ex}$1500$   &  \hspace*{-2ex}$20$    & \hspace*{-2ex}$90$   & \hspace*{-2ex}370 \\
\hspace*{-1.7ex}\citep{gabitov2015} & 3T GE & \hspace*{-2ex}axial &  \hspace*{-2ex}220  &   \hspace*{-2ex}$3.4\times3.4\times3.4$  & \hspace*{-2ex}$64\times64$ & \hspace*{-2ex}$3000$   &  \hspace*{-2ex}$35$    & \hspace*{-2ex}$90$   & \hspace*{-2ex}45 \\
\rowcolor{gray!13}
\hspace*{-1.7ex}\citep{lepping2016a} & 3T Siemens & \hspace*{-2ex}axial &  \hspace*{-2ex}220  &   \hspace*{-2ex}$2.9\times2.9\times3$  & \hspace*{-2ex}$64\times64$ & \hspace*{-2ex}$3000$   &  \hspace*{-2ex}$25$    & \hspace*{-2ex}$90$   & \hspace*{-2ex}105 \\
\rowcolor{gray!13}
	                  & Skyra (Erlangen, Germany) &           &                 &     &             &          &         &               &\\[1.5mm]
\hspace*{-1.7ex}\citep{iannilli2016} & 1.5T Siemens & \hspace*{-2ex}axial &  \hspace*{-2ex}-  &   \hspace*{-2ex}$3.3\times3.3\times3.3$  & \hspace*{-2ex}$64\times64$ & \hspace*{-2ex}$2500$   &  \hspace*{-2ex}$50$    & \hspace*{-2ex}-  & \hspace*{-2ex}120 \\
	                  & Aera (Erlangen, Germany) &           &                 &     &             &          &         &               &\\[1.5mm]
\rowcolor{gray!13}
\hspace*{-1.7ex}\citep{christian2017} & 1.5T GE Signa & \hspace*{-2ex}axial &  \hspace*{-2ex}$240$  &   \hspace*{-2ex}$4\times4\times4$  & \hspace*{-2ex}$64\times64$ & \hspace*{-2ex}$2000$   &  \hspace*{-2ex}$40$    & \hspace*{-2ex}90  & \hspace*{-2ex}267 \\
\hspace*{-1.7ex}\citep{kim2016} & 3T Siemens & \hspace*{-2ex}axial &  \hspace*{-2ex}-  &   \hspace*{-2ex}$3\times3\times3$  & \hspace*{-2ex}$64\times64$ & \hspace*{-2ex}$2200$   &  \hspace*{-2ex}$35$    & \hspace*{-2ex}90  & \hspace*{-2ex}365 \\
				  & Trio (Erlangen)    &           &                 &     &             &          &         &               &\\
\rowcolor{gray!13}
\hspace*{-1.7ex}\citep{romaniuk2016} & 3T Siemens & \hspace*{-2ex}axial & \hspace*{-2ex}220  &   \hspace*{-2ex}$3.4\times3.4\times5$  & \hspace*{-2ex}$64\times64$ & \hspace*{-2ex}$1560$   &  \hspace*{-2ex}$26$    & \hspace*{-2ex}66  & \hspace*{-2ex}341 \\
\rowcolor{gray!13}
				  & Magnetom Verio   &           &                 &     &             &          &         &               &\\
\hspace*{-1.7ex}\citep{arnab2017} & 3T Philips & \hspace*{-2ex}axial & \hspace*{-2ex}240  &   \hspace*{-2ex}$4\times4\times4$  & \hspace*{-2ex}$80\times80$ & \hspace*{-2ex}$2000$   &  \hspace*{-2ex}$30$    & \hspace*{-2ex}90  & \hspace*{-2ex}144 \\
				  & Achieva   &           &                 &     &             &          &         &               &\\
    \hline
\end{tabular}

    \caption{Data acquisition parameters for each fMRI study that we use
        for training DiFuMo atlases.
        Data are downloaded from OpenNeuro.}
    \label{tab:data_acquisition_somf_learner}
\end{table*}

\end{document}